\documentclass{iopart}
\usepackage[utf8]{inputenc}
\usepackage[T1]{fontenc}
\usepackage{enumitem}
\usepackage{cite}
\usepackage{amssymb}
\def\R{{\mathbb{R}}}

\begin{document}

\title[Cylindrical type integrable classical systems in a magnetic field]
{Cylindrical type integrable classical systems in a magnetic field}
\author{F Fournier$^1$, L {\v{S}}nobl$^2$  and P Winternitz$^3$}
\address{$^1$ Département de physique, Faculté des arts et des sciences, Universit\'e de Montr\'eal, CP 6128, Succ Centre-Ville, Montr\'eal (Qu\'ebec) H3C 3J7, Canada.}\ead{felix.fournier.1@umontreal.ca}
\address{$^2$ Department of Physics, Faculty of Nuclear Sciences and Physical Engineering, Czech Technical University in Prague, B\v rehov\'a 7, 115 19 Prague 1, Czech Republic}\ead{Libor.Snobl@fjfi.cvut.cz}
\address{$^3$
 Centre de recherches math\'ematiques and D\'epartement de math\'ematiques et de statistique, Universit\'e de Montr\'eal, CP 6128, Succ Centre-Ville, Montr\'eal (Qu\'ebec) H3C 3J7, Canada.}\ead{wintern@crm.umontreal.ca}
\begin{abstract}
We present all  second order classical integrable systems of the cylindrical type in a three dimensional Euclidean space $\mathbb{E}_3$ with a nontrivial magnetic field. The Hamiltonian and integrals of motion have the form 
\begin{eqnarray*}
\fl H =\frac{1}{2}\left(\vec{p}+\vec{A}(\vec{x})\right)^2+W(\vec{x}), \\
\fl  X_1=(p_\phi^A)^2+s_1^r(r, \phi, Z)p_r^A+s_1^\phi(r, \phi, Z)p_\phi^A+s_1^Z(r, \phi, Z)p_Z^A+m_1(r,\phi,Z), \nonumber \\
\fl X_2=(p_Z^A)^2+s_2^r(r, \phi, Z)p_r^A+s_2^\phi(r, \phi, Z)p_\phi^A+s_2^Z(r, \phi, Z)p_Z^A+m_2(r,\phi,Z).
\end{eqnarray*} Infinite families of such systems are found, in general depending on arbitrary functions or parameters. This leaves open the possibility of finding superintegrable systems among the integrable ones (i.e. systems with 1 or 2 additional independent integrals).\\

\end{abstract}
%\submitto{\jpa}

\section{Introduction}

This article is part of a research program the aim of which is to identify, classify and solve all superintegrable classical and quantum finite-dimensional Hamiltonian systems. We recall that a superintegrable system is one that allows more integrals of motion than degrees of freedom. For a review of the topic we refer to \cite{Miller_2013}. The best known superintegrable systems are given by the Kepler-Coulomb \cite{Fock_1935, Bargmann_1936, goldstein2002classical} and the harmonic oscillator potentials \cite{goldstein2002classical, Fradkin_1965, Jauch_1940}. A finite-dimensional classical Hamiltonian system in a $2n$-dimensional phase space is called integrable (or Liouville integrable) if it allows $n$ integrals of motion $\{ X_0=H,X_1,...,X_{n-1} \}$ (including the Hamiltonian $H$). These $n$ integrals must be well defined functions on the phase space. They must be in involution (Poisson commute pairwise, i.e. $\{ X_i, X_j \}_{P.B.} = 0$) and be functionally independent. The system is superintegrable if there exist further integrals $\{ Y_1,...,Y_k \}$, $1 \leq k \leq n-1$, that are also well defined functions on the phase space. The entire set $\{ X_0=H,X_1,\ldots,X_{n-1}, Y_1,\ldots,Y_k \}$ must be functionally independent and satisfy
\begin{eqnarray}
\{ H, X_j \}_{P.B.} = 0, \quad \{ X_i, X_j \}_{P.B.} = 0, \quad \{ H, Y_a \}_{P.B.} = 0, \\
i,j=1,\ldots,n-1, \quad a=1,\ldots,k, \quad 1 \leq k \leq n-1. \nonumber
\end{eqnarray}
Notice that $\{ Y_a, X_i \}_{P.B.} = 0$, $1 \leq i \leq n-1$, and $\{ Y_a, Y_b \}_{P.B.} = 0$ is not required. Moreover, the Poisson brackets $Z_{ai}=\{ Y_a, X_i \}_{P.B.}$ and $Z_{ab}=\{ Y_a, Y_b \}_{P.B.}$ generate a non--Abelian polynomial algebra.

A systematic search for ``natural'' Hamiltonians of the form
\begin{equation} \label{ham1}
H=\frac{1}{2}\vec{p}\;^2+W(\vec{x})
\end{equation} 
that are superintegrable in the $n$-dimensional Euclidean space $E_n$ started a long time ago \cite{Fris_1965, Winternitz_1967, Winternitz_1965, Makarov_1967} for $n=2$ and $n=3$. The integrals of motion $X_i$ and $Y_a$ were restricted to being second order polynomials in the components $p_i$ of the momenta. Second order integrals of motion were shown to be related to the separation of variables in the Hamilton-Jacobi equation (and also the Schrödinger equation). All second order superintegrable systems in $\mathbb{E}_2$ and $\mathbb{E}_3$ were found \cite{Fris_1965, Makarov_1967, Evans_1990}. Later developments for the Hamiltonian~\eref{ham1} and second order superintegrability include extensions to $\mathbb{E}_n$ for $n$ arbitrary, to general Riemannian, pseudo-Riemannian, and complex-Riemannian spaces \cite{MillerJr_1977, kalnins1986separation, Escobar_Ruiz_2017, Kalnins_2012, Kalnins_2007_1, Kalnins_2007_2, Kalnins_2003, Ranada_2017, Carinena_2017, Ballesteros_2011,RodWin,RodTemWin,Yehia,IvNerShma}.

More general Hamiltonians and their integrability and superintegrability properties are also being studied, in particular Hamiltonians with scalar and vector potentials both in $\mathbb{E}_2$ \cite{Dorizzi_1985, mcsween_2000, Berube_2004, Charest_2007, Pucacco_2005,Marikhin,MarSok,FerFor1,FerFor2} and $\mathbb{E}_3$ \cite{Zhalij_2015,Marchesiello_2015, Marchesiello_2017, Marchesiello_2018_1, Marchesiello_2018,Bertrand_2019,Shmavonyan}. 

In this article we focus on the case of a particle moving in an electromagnetic field in $\mathbb{E}_3$. It is described by a Hamiltonian with a scalar and vector potential, as in \cite{Marchesiello_2015, Marchesiello_2017, Marchesiello_2018_1, Marchesiello_2018, Bertrand_2019}. As opposed to previous articles, here we consider the ``cylindrical case'' when we have two second order integrals of motion of the ``cylindrical type''. In the absence of the vector potential the Hamiltonian would allow the separation of variables in cylindrical coordinates so that the potential in~\eref{ham1} would have the form 
\begin{equation}
W(\vec{r}) = W_1(r) + \frac{1}{r^2}W_2(\phi) + W_3(Z),
\end{equation} 
with the transformations $x=r \cos(\phi)$, $y=r \sin(\phi)$ and $z=Z$. 

\section{Formulation of the problem}

Let us consider a moving particle in an electromagnetic field, in a three-dimensional space. In cartesian coordinates, this simple system is described by the following Hamiltonian: 
\begin{equation}\label{HamMagn}
H=\frac{1}{2}\left(\vec{p}+\vec{A}(\vec{x})\right)^2+W(\vec{x})
\end{equation} 
where $\vec{p}=(p_1,p_2,p_3) \equiv (p_x,p_y,p_z)$ are the components of linear momentum, and $\vec{x}=(x_1,x_2,x_3) \equiv (x,y,z)$ are the cartesian spatial coordinates. The vector potential $\vec{A}(\vec{x})=\left(A_1(\vec{x}),A_2(\vec{x}),A_3(\vec{x})\right) \equiv \left(A_x(\vec{x}),A_y(\vec{x}),A_z(\vec{x})\right)$ and the scalar potential $W(\vec{x})$ depend only on the position $\vec{x}$. For practical reasons, the mass and electric charge of the particle have been set to $1$ and $-1$, respectively. 

The physical quantity related to the vector potential is the magnetic field 
\begin{equation}
\vec{B}(\vec{x})=\nabla \times \vec{A}(\vec{x}).
\end{equation} 
Let us consider integrals of motion which are at most quadratic in the momenta. They are of the form \cite{Marchesiello_2015}: 
\begin{equation}\label{IntCart}
\fl X=\sum_{j=1}^3 h^j(\vec{x})p_j^Ap_j^A + \sum_{j,k,l=1}^3 \frac{1}{2}|\epsilon_{jkl}|n^j(\vec{x})p_k^Ap_l^A + \sum_{j=1}^3 s^j(\vec{x})p_j^A + m(\vec{x})
\end{equation} 
where we have defined 
\begin{equation}
p_j^A=p_j+A_j(\vec{x})
\end{equation} 
and $h^j(\vec{x})$, $n^j(\vec{x})$, $s^j(\vec{x})$ ($j=1,2,3$) and $m(\vec{x})$ are real valued functions. They must satisfy the determining equations provided by the fact that the Poisson bracket of the integral with the Hamiltonian must vanish, i.e. 
\begin{equation}
\{H,X\}_{P.B.}=0
\end{equation} 
using the coefficients in front of each individual combination of powers in momenta. Those equations in cartesian coordinates are listed in previous papers \cite{Marchesiello_2015, Marchesiello_2017, Marchesiello_2018_1, Marchesiello_2018,Bertrand_2019}. 

It is possible to express the $h^j(\vec{x})$ and $n^j(\vec{x})$ functions as polynomials depending on $20$ real constants $\alpha_{ab}$, which allows us to say that the highest order terms of the integral $X$ are elements of the universal enveloping algebra of the Euclidean Lie algebra 
\begin{equation}
X=\sum_{1 \leq a \leq b \leq 6} \alpha_{ab}Y_a^A Y_b^A + \sum_{j=1}^3 s^j(\vec{x})p_j^A + m(\vec{x})
\end{equation} 
where 
\begin{equation}
Y^A=(p_1^A,p_2^A,p_3^A,l_1^A,l_2^A,l_3^A), \qquad l_i^A=\sum_{1 \leq j,k \leq 3} \epsilon_{ijk}x_j p_k^A.
\end{equation} 

We shall consider two integrals of motion $X_1$ and $X_2$ of the cylindrical type, in the sense that they imply separation of variables in cylindrical coordinates in the case of a vanishing magnetic field. Their exact form in the adequate system of coordinates will be specified below. 

We use the following relations between cartesian and cylindrical coordinates: 
\begin{equation}
x = r \cos(\phi), \qquad y = r \sin(\phi), \qquad z = Z.
\end{equation} 
Given the structure of the canonical 1-form 
\begin{equation}
\lambda = p_x \mathrm{d}x + p_y \mathrm{d}y + p_z \mathrm{d}z = p_r \mathrm{d}r + p_\phi \mathrm{d}\phi + p_Z \mathrm{d}Z,
\end{equation} 
we obtain the following transformation for the linear momentum: 
\begin{eqnarray}
\fl  p_x &= \cos(\phi)p_r-\frac{\sin(\phi)}{r}p_\phi, \quad   p_y = \sin(\phi)p_r+\frac{\cos(\phi)}{r}p_\phi,  \quad  p_z = p_Z
\end{eqnarray}
and similarly for the components of the vector potential. On the other hand, the components of the magnetic field are the components of the 2-form $B=\mathrm{d}A$,
\begin{eqnarray}
\fl  B &= B^x (\vec{x})\mathrm{d}y \wedge \mathrm{d}z + B^y (\vec{x})\mathrm{d}z \wedge \mathrm{d}x + B^z (\vec{x})\mathrm{d}x \wedge \mathrm{d}y \nonumber \\
\fl  &= B^r (r,\phi,Z)\mathrm{d}\phi \wedge \mathrm{d}Z + B^\phi (r,\phi,Z)\mathrm{d}Z \wedge \mathrm{d}r + B^Z(r,\phi,Z)\mathrm{d}r \wedge \mathrm{d}\phi.
\end{eqnarray}
This leads to the following transformation
\begin{eqnarray} \label{transformB}
   B^x (\vec{x}) &= \frac{\cos(\phi)}{r}B^r (r,\phi,Z) - \sin(\phi)B^\phi (r,\phi,Z), \nonumber \\
  B^y (\vec{x}) &= \frac{\sin(\phi)}{r}B^r (r,\phi,Z) + \cos(\phi)B^\phi (r,\phi,Z), \\
  B^z (\vec{x}) &= \frac{1}{r}B^Z (r,\phi,Z). \nonumber
\end{eqnarray} 

We can now rewrite both the Hamiltonian and the general form of an integral of motion in cylindrical coordinates. 

\section{Hamiltonian and integrals of motion in the cylindrical case}\label{Himcc}

We will first write down the general form of the Hamiltonian and integrals in cylindrical coordinates, and then restrict to the case of two integrals of motion which correspond to the so-called cylindrical case. 

\subsection{Determining equations in cylindrical coordinates}

In cylindrical coordinates, the Hamiltonian~\eref{HamMagn} takes the following form: 
\begin{equation}\label{HamMagnCyl}
 H=\frac{1}{2}\left(\left(p_r^A\right)^2+\frac{\left(p_\phi^A\right)^2}{r^2}+\left(p_Z^A\right)^2\right)+W(r,\phi,Z),
\end{equation}
where
\begin{equation}
\fl p_r^A = p_r + A_r (r,\phi,Z), \quad p_\phi^A = p_\phi + A_\phi (r,\phi,Z), \quad p_Z^A = p_Z + A_Z (r,\phi,Z).
\end{equation} 
The integral of motion~\eref{IntCart} now reads as follows 
\begin{eqnarray}\label{IntCyl}
  X &= h^r \left(r,\phi,Z\right)\left(p_r^A\right)^2 + h^\phi \left(r,\phi,Z\right)\left(p_\phi^A\right)^2 + h^Z \left(r,\phi,Z\right)\left(p_Z^A\right)^2 + \nonumber \\
  &+ n^r \left(r,\phi,Z\right)p_\phi^A p_Z^A + n^\phi \left(r,\phi,Z\right)p_r^A p_Z^A + n^Z \left(r,\phi,Z\right)p_\phi^A p_r^A + \\
  &+ s^r \left(r,\phi,Z\right)p_r^A + s^\phi \left(r,\phi,Z\right)p_\phi^A + s^Z \left(r,\phi,Z\right)p_Z^A + m\left(r,\phi,Z\right). \nonumber\end{eqnarray} 
The functions $h^r$, ..., $n^Z$ can be obtained from the $h^j$ and $n^j$ via their transformations into cylindrical coordinates, and are expressed in terms of the same $20$ constants $\alpha_{ab}$. 

Computing the Poisson bracket $\{H,X\}_{P.B.}$ in the cylindrical coordinates we obtain terms of order 3, 2, 1 and 0 in the components of $\vec{p}^A$. The third order terms provide the following determining equations
\begin{eqnarray} \label{ord3}
\fl  \partial_r h^r &= 0, \quad \partial_\phi h^r = -r^2 \partial_r n^Z, \quad \partial_Z h^r = - \partial_r n^\phi, \nonumber \\
\fl  \partial_r h^\phi &= -\frac{1}{r^2}\partial_\phi n^Z - \frac{2}{r^3}h^r, \quad \partial_\phi h^\phi = -\frac{1}{r}n^Z, & \partial_Z h^\phi = -\frac{1}{r^2}\partial_\phi n^r - \frac{1}{r^3}n^\phi, \nonumber \\
\fl  \partial_r h^Z &= -\partial_Z n^\phi, \quad \partial_\phi h^Z = -r^2 \partial_Z n^r, \quad \partial_Z h^Z = 0,  \\ \nonumber
\fl   \partial_\phi n^\phi &= -r^2(\partial_Z n^Z + \partial_r n^r).
\end{eqnarray}
In the second order terms we use equations~\eref{ord3} and rewrite derivatives of the vector potential $\vec A$ in terms of the magnetic field $\vec B$, to obtain 
\begin{eqnarray} \label{ord2}
  \partial_r s^r &= n^\phi B^\phi - n^Z B^Z, \nonumber \\
  \partial_\phi s^r &= r^2(n^r B^\phi - 2h^\phi B^Z - \partial_r s^\phi) - n^\phi B^r + 2h^r B^Z, \nonumber \\
  \partial_r s^Z &= n^Z B^r - \partial_Z s^r - n^r B^Z + 2h^Z B^\phi - 2h^r B^\phi, \nonumber \\
  \partial_\phi s^\phi &= -n^r B^r + n^Z B^Z - \frac{1}{r}s^r, \\
  \partial_\phi s^Z &= r^2(2h^\phi B^r - n^Z B^\phi - \partial_Z s^\phi) - 2h^Z B^r + n^\phi B^Z, \nonumber \\
  \partial_Z s^Z &= n^r B^r - n^\phi B^\phi. \nonumber
\end{eqnarray} 
The first and zeroth order terms imply 
\begin{eqnarray} \label{ord1}
  \partial_r m &= s^Z B^\phi - s^\phi B^Z + n^\phi \partial_Z W + n^Z \partial_\phi W + 2h^r \partial_r W, \nonumber \\
  \partial_\phi m &= s^r B^Z - s^Z B^r + r^2(n^r \partial_Z W + 2h^\phi \partial_\phi W + n^Z \partial_r W), \\
  \partial_Z m &= s^\phi B^r - s^r B^\phi + 2h^Z \partial_Z W + n^r \partial_\phi W + n^\phi \partial_r W, \nonumber
\end{eqnarray}
and
\begin{equation}\label{ord0}
	s^r \partial_r W + s^\phi \partial_\phi W + s^Z \partial_Z W = 0,
\end{equation}
respectively. 

\subsection{Reduction to the cylindrical case}

The integrals of motion corresponding to the cylindrical case, i.e. the case which allows separation of variables in cylindrical coordinates for a vanishing magnetic field, read
\begin{eqnarray} \label{cylintegrals} 
\fl X_1=(p_\phi^A)^2+s_1^r(r, \phi, Z)p_r^A+s_1^\phi(r, \phi, Z)p_\phi^A+s_1^Z(r, \phi, Z)p_Z^A+m_1(r,\phi,Z), \nonumber \\
\fl X_2=(p_Z^A)^2+s_2^r(r, \phi, Z)p_r^A+s_2^\phi(r, \phi, Z)p_\phi^A+s_2^Z(r, \phi, Z)p_Z^A+m_2(r,\phi,Z).
\end{eqnarray}
For such integrals with specific values for the $h$ and $n$ coefficients, all of them being either $0$ or $1$, it follows that system~\eref{ord3} is satisfied trivially for both $X_1$ and $X_2$. The system~\eref{ord2} applied to both integrals gives the following equations: 
\begin{eqnarray} \label{cyl2a}
\partial_r s_1^r &= 0, \quad \partial_\phi s_1^\phi = -\frac{s_1^r}{r}, \nonumber \\
\partial_\phi s_1^r &= -r^2(\partial_r s_1^\phi + 2 B^Z), \quad \partial_\phi s_1^Z = r^2(-\partial_Z s_1^\phi + 2 B^r), \\
\partial_r s_1^Z &= -\partial_Z s_1^r, \quad \partial_Z s_1^Z = 0, \nonumber \end{eqnarray}
\begin{eqnarray}
\partial_r s_2^r &= 0, \quad \partial_\phi s_2^\phi = -\frac{s_2^r}{r}, \nonumber \\ \label{cyl2b}
\partial_\phi s_2^r &= -r^2 \partial_r s_2^\phi, \quad \partial_\phi s_2^Z = -r^2 \partial_Z s_2^\phi - 2 B^r,\\
\partial_r s_2^Z &= -\partial_Z s_2^r + 2 B^\phi, \quad \partial_Z s_2^Z = 0. \nonumber
\end{eqnarray} 
The systems~\eref{ord1} and~\eref{ord0} reduce to 
\begin{eqnarray} \label{cyl1a}
\partial_r m_1 &= s_1^Z B^\phi - s_1^\phi B^Z, \nonumber \\
\partial_\phi m_1 &= s_1^r B^Z - s_1^Z B^r + 2 r^2 \partial_\phi W,\\
\partial_Z m_1 &= s_1^\phi B^r - s_1^r B^\phi, \nonumber 
\end{eqnarray}
\begin{eqnarray}
\partial_r m_2 &= s_2^Z B^\phi - s_2^\phi B^Z, \nonumber \\ \label{cyl1b}
\partial_\phi m_2 &= s_2^r B^Z - s_2^Z B^r,\\
\partial_Z m_2 &= s_2^\phi B^r - s_2^r B^\phi + 2 \partial_Z W, \nonumber
\end{eqnarray}
and
\begin{equation} \label{cyl0}
s_i^r \partial_r W + s_i^\phi \partial_\phi W + s_i^Z \partial_Z W = 0 \qquad (i=1,2),
\end{equation} 
respectively.

Let us now consider the Poisson bracket $\{X_1,X_2\}_{P.B.}$, which must also vanish for an integrable system. This provides further equations for every order in the momenta. First, for the second order, we have 
\begin{eqnarray} \label{extra2}
\partial_\phi s_2^\phi &= 0, \quad \partial_\phi s_2^r = 0, \quad \partial_Z s_1^r = 0, \quad \partial_\phi s_2^Z = \partial_Z s_1^\phi - 2 B^r.
\end{eqnarray} 
From those, we can already conclude, looking again at system~\eref{cyl2b}, that $s_2^r = 0$. The first order terms in the same Poisson bracket $\{X_1,X_2\}_{P.B.}$ imply
\begin{eqnarray} \label{extra1}
s_2^Z \partial_Z s_1^r + s_2^\phi \partial_\phi s_1^r &= 0, \nonumber \\
-s_1^\phi(2 B^r + \partial_\phi s_2^Z) + s_2^Z \partial_Z s_1^Z - s_1^Z \partial_Z s_2^Z \nonumber \\
+ s_2^\phi \partial_\phi 2_1^Z + s_1^r(2 B^\phi - \partial_r s_2^Z) + 2 \partial_Z m_1 &= 0,\\
-s_2^Z(2 B^r - \partial_Z s_1^\phi) + s_2^\phi \partial_\phi s_1^\phi \nonumber \\
- s_1^Z \partial_Z s_2^\phi - s_1^r \partial_r s_2^\phi - 2 \partial_\phi m_2 &= 0. \nonumber 
\end{eqnarray} 
From the zeroth order term we obtain 
\begin{eqnarray} \label{extra0}
- s_1^r \partial_r m_2 + s_2^\phi \partial_\phi m_1 - s_1^\phi \partial_\phi m_2 + s_2^Z \partial_Z m_1 - s_1^Z \partial_Z m_2 \nonumber \\
+ B^r(s_2^\phi s_1^Z - s_1^\phi s_2^Z) + B^\phi s_1^r s_2^Z - B^Z s_1^r s_2^\phi &= 0.
\end{eqnarray} 

\section{Partial solution of determining equations and reduction to functions of one variable}\label{Psderfov}

The second order terms in momenta from the aforementioned vanishing Poisson brackets, i.e. systems~\eref{cyl2a},~\eref{cyl2b} and~\eref{extra2}, provide a system of equations for the functions $s_j^{r,\phi,Z}$ and the magnetic field $B^{r,\phi,Z}$ which can be easily solved. The solution is expressed in terms of $5$ functions of one variable each: $\sigma(r)$, $\rho(r)$, $\tau(\phi)$, $\psi(\phi)$ and $\mu(Z)$. We shall call them the auxiliary functions: 
\begin{eqnarray} \label{scond}
s_1^r &=\frac{\mathrm{d}}{\mathrm{d}\phi}\psi(\phi), \quad s_1^\phi=-\frac{\psi(\phi)}{r}-r^2\mu(Z)+\rho(r), \quad s_1^Z=\tau(\phi), \nonumber \\
s_2^r&=0, \quad s_2^\phi=\mu(Z), \quad s_2^Z=-\frac{\tau(\phi)}{r^2}+\sigma(r)
\end{eqnarray} 
\begin{eqnarray} \label{Bcond}
B^r&=-\frac{r^2}{2}\frac{\mathrm{d}}{\mathrm{d}Z}\mu(Z)+\frac{1}{2r^2}\frac{\mathrm{d}}{\mathrm{d}\phi}\tau(\phi), \quad B^\phi=\frac{\tau(\phi)}{r^3}+\frac{1}{2}\frac{\mathrm{d}}{\mathrm{d}r}\sigma(r), \nonumber \\
B^Z&=\frac{-\psi(\phi)}{2r^2}+r\mu(Z)-\frac{1}{2}\frac{\mathrm{d}}{\mathrm{d}r}\rho(r)-\frac{1}{2r^2}\frac{\mathrm{d}^2}{\mathrm{d}\phi^2}\psi(\phi).
\end{eqnarray} 
Equations~\eref{scond} and~\eref{Bcond} are the general solutions of equations~\eref{cyl2a},~\eref{cyl2b} and~\eref{extra2}. We use them to eliminate the functions $\vec{s}_1,\vec{s}_2$ and $\vec{B}$ from the as yet unsolved PDEs~(\ref{cyl1a}-\ref{cyl0}) and~(\ref{extra1}-\ref{extra0}).
Using~(\ref{cyl1a}-\ref{cyl1b}) and~\eref{extra1} we end up with one equation for each possible first derivative of both $m_1$ and $m_2$, one direct condition on $\mu(Z)$ and $\psi(\phi)$, and two equations which are further conditions on $m_{1,Z}$ and $m_{2,\phi}$
\begin{eqnarray} \label{1_0cond}
\fl \left( -r^3 \mu(Z) + r \rho(r) - \psi(\phi) \right) \left( \psi''(\phi) + r^2 \rho'(r) \right) + \left( r^3 \mu(Z) + r \rho(r) \right) \psi(\phi) \nonumber \\
\fl - \psi(\phi)^2 + r^3 \tau(\phi) \sigma'(r) + 2 r^6 \mu(Z)^2 - 2r^4 \rho(r) \mu(Z) + 2 \tau(\phi)^2 - 2r^3 m_{1,r} &=0, \nonumber \\
\fl \psi'(\phi) \left( 2r^3 \mu(Z) - r^2 \rho'(r) - \psi(\phi) - \psi''(\phi) \right) \nonumber \\
\fl + \tau(\phi) \left( r^4 \mu'(Z) - \tau'(\phi) \right) + 4r^4 W_\phi - 2r^2 m_{1,\phi} &= 0, \nonumber \\
\fl \left( \tau'(\phi) - r^4 \mu'(Z) \right) \left( -r^3 \mu(Z) + r \rho(r) - \psi(\phi) \right) \nonumber \\
\fl - \psi'(\phi) \left( r^3 \sigma'(r) + 2\tau(\phi) \right) -2r^3 m_{1,Z} &= 0, \nonumber \\
\fl r^3 \mu(Z) \psi''(\phi) + r^3 \sigma'(r) \left( r^2 \sigma(r) - \tau(\phi) \right) - 2r^6 \mu(Z)^2 \nonumber \\
\fl + r^5 \mu(Z) \rho'(r) + r^3 \mu(Z) \psi(\phi) + 2r^2 \sigma(r) \tau(\phi) - 2\tau(\phi)^2 - 2r^5 m_{2,r} &= 0, \nonumber \\
\fl \left( r^4 \mu'(Z) - \tau'(\phi) \right) \left( r^2 \sigma(r) - \tau(\phi) \right) -2r^4 m_{2,\phi} &=0,  \\
\fl -r^4 \mu(Z) \mu'(Z) + \mu(Z) \tau'(\phi) + 4r^2 W_Z - 2r^2 m_{2,Z} &=0, \nonumber \\
\fl \mu(Z)\psi''(\phi)&=0, \nonumber \\
\fl  \left( -r^4 \mu(Z) + r^2 \rho(r) - r \psi(\phi) \right) \mu'(Z) + \mu(Z) \tau'(\phi) + 2 m_{1,Z} &= 0, \nonumber \\
\fl \tau'(\phi) \left( r^2 \sigma(r) - \tau(\phi) \right) + r^4 \tau(\phi) \mu'(Z) + r^3 \mu(Z) \psi'(\phi) + 2r^4 m_{2,\phi} &= 0. \nonumber \\
\fl \,\nonumber
\end{eqnarray}

From~\eref{cyl0} and~\eref{extra0} we obtain 3 further equations

\begin{eqnarray} \label{0_0cond}
\fl \left(r^2 \sigma(r) - \tau(\phi)\right)W_Z + r^2 \mu(Z) W_\phi &= 0, \nonumber \\
\fl \left(-r^3 \mu(Z) + r \rho(r) - \psi(\phi)\right)W_\phi + r(\psi'(\phi)W_r + \tau(\phi)W_Z) &= 0, \nonumber \\
\fl 2r^4 \left( r^3 \mu(Z) - r \rho(r) + \psi(\phi) \right) m_{2,\phi} + 2 r^3 \left( r^2 \sigma(r) - \tau(\phi) \right) m_{1,Z} \nonumber \\
\fl + r^3 \mu(Z) \psi'(\phi)\psi''(\phi) + 2r^5 \mu(Z) m_{1,\phi} - 2r^5 \tau(\phi) m_{2,Z} \nonumber \\
\fl + \left( r^3 \left( r^2 \sigma(r) - \tau(\phi) \right)\sigma'(r) -2r^6 \mu(Z)^2 +r^5 \mu(Z) \rho'(r) \right.\\
\fl \left. + r^3 \mu(Z) \psi(\phi) + 2r^2 \sigma(r) \tau(\phi) -2\tau(\phi)^2 - 2r^5 m_{2,r} \right)\psi'(\phi) \nonumber \\
\fl - \left( r^2 \left( r^3 \mu(Z) - r \rho(r) + \psi(\phi) \right)\sigma(r) \right. \nonumber \\
\fl + \left. \tau(\phi) \left( r \rho(r) - \psi(\phi) \right) \right) \left( r^4 \mu'(Z) - \tau'(\phi) \right) &= 0. \nonumber 
\end{eqnarray} 

Before summing up the results of this section in the form of a reduced system of determining equations let us analyze the PDEs~\eref{1_0cond} and~\eref{0_0cond}. First of all, $m_{1,Z}$ and $m_{2,\phi}$ appear in~\eref{1_0cond} twice each. Since the two values must coincide we obtain two constraints on the auxiliary functions. A further constraint $\mu(Z)\psi''(\phi)=0$ is explicit in~\eref{1_0cond}. The remaining 6 equations in~\eref{1_0cond} are used to express all first order derivatives $m_{1,a}$, $m_{2,a}$ ($a=r,\phi,Z$) in terms of $W_\phi$, $W_Z$ and the auxiliary functions. Assuming that the functions $m_1$ and $m_2$ are sufficiently smooth we impose the Clairaut compatibility conditions $\partial_a\partial_b m_{i}=\partial_b\partial_a m_{i}$ on their second derivatives. This gives us a further set of equations
\begin{eqnarray} \label{compcond2}
\fl m_{1,r \phi}: &\quad \psi'(\phi)\left(-3 \psi''(\phi) + r^3 \rho''(r) - r^3 \mu(Z) - r^2 \rho'(r) + r \rho(r) - 4 \psi(\phi)\right) & \nonumber \\
\fl &+ \tau'\left(\phi)(r^3 \sigma'(r) + 2 \tau(\phi)\right) - 2 r^4 \tau(\phi) \mu'(Z) - 4r^5 W_{r\phi} - 8 r^4 W_\phi \nonumber & \\
\fl &+ \left(r \rho(r) - \psi(\phi)\right)\psi'''(\phi) & = 0, \nonumber \\
\fl m_{1,r Z}: &\quad -r^4 \mu'(Z)\psi''(\phi) + r^4 \psi'(\phi)\sigma''(r) -6 \tau(\phi) \psi'(\phi) & \nonumber \\
\fl &+ \tau'(\phi) \left( -r^2 \rho'(r) + 2r \rho(r) - 3\psi(\phi) \right) & = 0, \nonumber \\
\fl m_{1,\phi Z}: &\quad \tau''(\phi) \left( r^3 \mu(Z) - r \rho(r) + \psi(\phi) \right) + \psi''(\phi) \left( r^3 \sigma'(r)+ 2 \tau(\phi) \right) & \nonumber \\
\fl & + r^5 \tau(\phi) \mu''(Z) + \psi'(\phi) \left( r^4 \mu'(Z) + 3 \tau'(\phi) \right) + 4r^5 W_{\phi Z} & = 0, \\
\fl m_{2,r \phi}: &\quad -r^3 \mu'(Z)\left(r \sigma'(r) + 2 \sigma(r)\right) + \mu(Z)\psi'(\phi) & = 0, \nonumber \\
\fl m_{2,r Z}: &\quad r\mu'(Z)\left( - 2r^3 \mu(Z) + r^2 \rho'(r) + \psi(\phi)\right) -4r^3 W_{rZ} + 2 \mu(Z) \tau'(\phi) & = 0, \nonumber \\
\fl m_{2,\phi Z}: &\quad r^2\left(\tau(\phi)-r^2 \sigma(r)\right)\mu''(Z) + \tau''(\phi)\mu(Z) + 4 r^2 W_{\phi Z} & = 0. \nonumber
\end{eqnarray}
Equations~\eref{compcond2} can be solved for the second mixed derivatives of the potential $W_{r\phi}$, $W_{rZ}$ and $W_{\phi Z}$ in terms of $W_{\phi}$ and the auxiliary functions. The identities for the mixed third order derivatives of $W$ are satisfied identically as a consequence of the compatibility of the second order ones.

Finally we substitute the first order derivatives  $m_{1,a}$, $m_{2,a}$ from~\eref{1_0cond} into~\eref{0_0cond} and obtain a system of linear inhomogeneous algebraic equations for the first order derivatives $W_r$, $W_\phi$, $W_Z$. Implementing the procedure described above we obtain the reduced system of determining equations presented in the following Section~\ref{reddetsys}.

\section{Reduced determining system}\label{reddetsys}

The determining system now reduces to two conditions on the auxiliary functions, three equations from~\eref{compcond2} that involve mixed second derivatives of $W$, and a linear algebraic system involving all first derivatives of $W$. We list them all here:
\numparts
\begin{eqnarray} 
\psi'(\phi) \left( r^3 \sigma'(r) + 2 \tau(\phi) \right) - \tau'(\phi) \left( r \rho(r) - \psi(\phi) \right) = 0, \label{reducedAa} \\
\mu(Z) \psi'(\phi) + r^3 \sigma(r)\mu'(Z) = 0, \label{reducedAb}
\end{eqnarray}
\endnumparts
\begin{eqnarray} \label{reducedB}
\fl W_{r \phi} &= -\frac{2}{r}W_\phi + \frac{1}{4r^5}\left( \psi'(\phi) \left( -3\phi''(\phi) + r^3 \rho''(r) - r^3 \mu(Z) - r^2 \rho'(r) + r \rho(r) - 4\psi(\phi) \right) \right. \nonumber \\
\fl &+ \left. \tau'(\phi) \left( r^3 \sigma'(r) + 2\tau(\phi) \right) - 2r^4 \tau(\phi) \mu'(Z) - \psi'''(\phi) \left( \psi(\phi) - r \rho(r) \right) \right), \nonumber \\
\fl W_{\phi Z} &= -\frac{1}{4r^2} \left( r^2 \mu''(Z) \left( \tau(\phi) - r^2 \sigma(r) \right) + \tau''(\phi) \mu(Z) \right), \\
\fl W_{r Z} &= \frac{1}{4r^3} \left( r \mu'(Z) \left( -2r^3 \mu(Z) + r^2 \rho'(r) + \psi(\phi) \right) + 2 \mu(Z) \tau'(\phi) \right), \nonumber
\end{eqnarray}
\begin{eqnarray} \label{matrixform2}
\fl {\underbrace{
\left( \begin{array}{ccc}
0 & r^2 \mu(Z) &  r^2 \sigma(r) - \tau(\phi) \\
\psi'(\phi) & \rho(r) -r^2 \mu(Z) - \frac{\psi(\phi)}{r} & \tau(\phi)\\
0 & 4 r^7 \mu(Z) & -4 r^5 \tau(\phi)
\end{array}\right)
}} 
 &
\cdot  \underbrace{
\left(\begin{array}{c}
W_r\\
W_\phi\\
W_Z
\end{array}\right)}
& = 
\underbrace{\left(\begin{array}{c}
0\\
0\\
\alpha(r,\phi,Z)
\end{array}\right)} \nonumber \\ 
\qquad \quad  M  & \cdot \quad \nabla W & = \qquad \;\, \vec\alpha
\end{eqnarray} 
where 
\begin{eqnarray} \label{alpha}
\fl \alpha(r,\phi,Z)  = & -\psi'(\phi)
 \left( \left(-r^5 \sigma(r) + r^3 \tau(\phi)\right)\sigma'(r) - r^5 \mu(Z) \rho'(r) + 2 \tau(\phi)^2 \nonumber  \right. \\
\fl & \left. - 2 r^2 \sigma(r) \tau(\phi) + r^3 \mu(Z)\left(r^3 \mu(Z) + r \rho(r) - 2 \psi(\phi)\right)\right) 
 \nonumber \\
\fl &- \tau'(\phi)\left(\left(-r \rho(r) + \psi(\phi)\right)\tau(\phi) - r^2 \sigma(r)\left(r^3 \mu(Z) - r \rho(r) + \psi(\phi)\right)\right) \nonumber \\
\fl &- r^4 \mu'(Z) \tau(\phi)\left(r \rho(r) - \psi(\phi)\right).
\end{eqnarray} 
The rank of the matrix $M$ can be either $3$, $2$ or $1$. We rule out the rank $0$ case since it leads to vanishing magnetic field, as seen directly from~\eref{Bcond}. 

If the rank is $3$, then the determinant of $M$
\begin{equation} \label{detM}
\det(M) = 4 r^{9} \psi'(\phi)\mu(Z)\sigma(r).
\end{equation} 
 is not zero and it implies a unique solution for each first derivative of $W$. We will explore this case shortly and show that it leads to a contradiction. 

If instead the rank is either $2$ or $1$, then $\det(M)=0$, and from~\eref{detM}, there are {\it a priori} three possible cases:
\begin{enumerate}[label=\alph*)]
\item $\psi'(\phi)=0$,
\item $\psi'(\phi) \neq 0$ and $\mu(Z)=0$,
\item $\psi'(\phi) \neq 0$, $\mu(Z) \neq 0$ and $\sigma(r)=0$. However, we observe that this is inconsistent with~\eref{reducedAb}, so we can already rule this case out.
\end{enumerate}

We shall first show that we must have $\alpha=0$ in all these cases, allowing us to simplify further considerations below. 
\begin{enumerate}[label=\alph*)]
\item $\psi'(\phi)=0$. This is equivalent to $\psi(\phi)=0$ since the function $\psi$ has to be constant and thus it can be absorbed into a redefinition of $\rho(r)$ in equations~\eref{scond} and~\eref{Bcond}. The augmented matrix of the system of linear equations~\eref{matrixform2} can be written in its reduced row echelon form as
\begin{equation}
\left(\begin{array}{cccc}
0 & r^2 \mu(Z) & -\tau(\phi) & \frac{\alpha}{4r^5}\\
0 & \rho(r) & 0 & \frac{\alpha}{4r^5}\\
0 & 0 & \sigma(r) & -\frac{\alpha}{4r^7}
\end{array}\right)
\end{equation} 
From~(\ref{reducedAa}-\ref{reducedAb}) we have 
\begin{equation}\label{reducedApsi0}
\tau'(\phi)\rho(r)=0, \quad \mu'(Z)\sigma(r)=0.
\end{equation}
Consequently, the expression for $\alpha$ reads
\begin{equation}
\alpha = r^5 \left( \tau'(\phi)\sigma(r)\mu(Z) - \mu'(Z)\rho(r)\tau(\phi) \right).
\end{equation}
Equations~\eref{reducedApsi0} give rise to four possible solutions
\begin{itemize}
\item $\tau'(\phi)=0, \, \mu'(Z)=0$, implying $\alpha=0$ directly,
\item $\rho(r)=0, \, \sigma(r)=0$, implying $\alpha=0$ directly,
\item $\tau'(\phi)=0, \, \sigma(r)=0$, implying $\alpha = - r^5 \mu'(Z)\rho(r)\tau(\phi)$,
\item $\rho(r)=0, \, \mu'(Z)=0$, implying $\alpha = r^5 \tau'(\phi)\sigma(r)\mu(Z)$.
\end{itemize}
On the other hand, the solvability condition of the linear system~\eref{matrixform2}, namely that the rank of $M$ and of the corresponding augmented matrix coincide, imply that if either $\rho(r)=0$ or $\sigma(r)=0$, the function $\alpha$ must vanish. Thus in the two cases above we find constraints,
\begin{itemize}
\item if $\psi'(\phi)=\tau'(\phi)=\sigma(r)=0$ we must have 
\begin{equation}\label{alpha0a}
\mu'(Z)\rho(r)\tau(\phi)=0,
\end{equation}
\item if $\psi'(\phi)=\rho(r)=\mu'(Z)=0$ we must have 
\begin{equation}\label{alpha0b}
\tau'(\phi)\sigma(r)\mu(Z)=0.
\end{equation}
\end{itemize} 
\item $\psi'(\phi) \neq 0$ and $\mu(Z)=0$. In this case equation~\eref{reducedAb} is satisfied trivially. Equation~\eref{reducedAa} we differentiate with respect to $r$, arriving at
\begin{equation}\label{reducedAadr}
\left( r^3 \sigma'(r)\right)' = \frac{\tau'(\phi)}{\psi'(\phi)} \left( r \rho(r)\right)',
\end{equation}
leading to three distinct possibilities
\begin{itemize}
\item $\left( r^3 \sigma'(r)\right)' =\left( r \rho(r)\right)'=0$, i.e.
\begin{equation}\label{reducedAadr1}
\sigma(r)=\frac{C_\sigma}{r^2}+\tilde{C}_\sigma, \qquad \rho(r)= \frac{C_\rho}{r}.
\end{equation}
Substituting~\eref{reducedAadr1} into equation~\eref{reducedAa} we find 
\begin{equation}
2 \left( \tau(\phi) -C_{\sigma}\right) \psi'(\phi)+ \left(\psi(\phi)-C_{\rho}\right ) \tau'(\phi)=0
\end{equation}
which directly implies that $\alpha$ defined in~\eref{alpha} vanishes.
\item $\left( r^3 \sigma'(r)\right)' =\tau'(\phi)=0$, i.e.
\begin{equation}\label{reducedAadr2}
\sigma(r)=\frac{C_{\sigma}}{r^2}+\tilde{C}_{\sigma}, \qquad \tau(\phi)=C_{\tau}.
\end{equation}
Substituting~\eref{reducedAadr2} into equation~\eref{reducedAa} we find $C_{\sigma}=C_{\tau}$ and that together with equation~\eref{reducedAadr2} implies again that we find $\alpha=0$ in~\eref{alpha}.
\item $\frac{\left( r^3 \sigma'(r)\right)'}{\left( r \rho(r)\right)'} =  \frac{\tau'(\phi)}{\psi'(\phi)} = \lambda \neq 0,$ implying that
\begin{equation}\label{reducedAadr3}
\rho(r)=\frac{1}{\lambda} r^2 \sigma'(r)+\frac{C_\rho}{r}, \qquad \tau(\phi)= \lambda \psi(\phi)+C_\tau.
\end{equation}
However, substituting~\eref{reducedAadr3} into~\eref{reducedAa} and differentiating it with respect to $\phi$ we arrive at $\lambda \psi'(\phi) =0$ which contradicts our assumptions $\lambda\neq 0$ and $ \psi'(\phi)\neq 0$.
\end{itemize}
\end{enumerate}
Thus we see that for all solutions of the determining equations we have $\alpha=0$. In most cases $\alpha=0$ by virtue of~(\ref{reducedAa}-\ref{reducedAb}) alone, in two cases the condition that the augmented matrix of the system~\eref{matrixform2} and the matrix $M$ have the same rank leads to certain additional constraints, cf. equations~\eref{alpha0a} and~\eref{alpha0b}.\medskip

We are now ready to split the classification problem into three main cases according to the rank of the matrix $M$, which leads to various classes of potentials and magnetic fields.

\section{Solutions of determining equations for Case 1: $\det(M) \neq 0$ ($\mathrm{rank}(M) = 3$)}\label{SdeM3}

Let's begin with the seemingly most complicated case: the case where the determinant of $M$ is not equal to zero, or in other words, the rank of $M$ is $3$. We are going to prove that this case leads to an inconsistency and has no solutions. 

Recalling~\eref{detM}, this requires that $\psi'(\phi) \neq 0, \mu(Z) \neq 0$ and $\sigma(r) \neq 0$. From~\eref{reducedAa} we have $\psi''(\phi)=0$. We can assume that $\psi(\phi) = \psi_1 \phi$ where the constant $\psi_1$ satisfies $\psi_1\neq 0$, since an additive constant would be absorbed into $\rho(r)$ by a simple redefinition. Looking at equation~\eref{reducedAb}, it becomes obvious that $\sigma(r)$ takes the following form
\begin{equation}
\sigma(r) = \frac{\sigma_0}{r^3}, \quad \sigma_0 \neq 0.
\end{equation} 
Equation~\eref{reducedAa} then becomes 
\begin{equation} \label{sigcond}
\psi_1 \left( -3 \sigma_0 + 2 r \tau(\phi) \right) - \tau'(\phi) \left( r^2 \rho(r) - r \psi_1 \phi \right) = 0.
\end{equation} 
Differentiation with respect to $r$ gives 
\begin{equation} \label{derived}
2 \psi_1 \tau(\phi) - \tau'(\phi) \left( 2 r \rho(r) + r^2 \rho'(r) - \psi_1 \phi \right) = 0.
\end{equation} 
From this point we can separate the variables $r$ and $\phi$ if $\tau'(\phi) \neq 0$. Notice that this has to be true since $\tau'(\phi) = 0$ would imply that either $\psi_1$ or $\tau(\phi)$ is zero, from the previous equation. The latter is not possible in view of~\eref{sigcond} since it would imply that $\sigma_0=0$, which contradicts our initial hypothesis. This means that we can rewrite~\eref{derived} as 
\begin{equation}
2 r \rho(r) + r^2 \rho'(r) = k = \psi_1 \phi + \frac{2 \psi_1 \tau(\phi)}{\tau'(\phi)},
\end{equation} 
where $k$ is a constant. Solving for $\rho(r)$, we have
\begin{equation}
\rho(r) = \frac{\rho_0}{r^2} + \frac{k}{r}.
\end{equation} 
Heading back to~\eref{reducedAa} using the newly known expression for $\rho(r)$ and separating the various powers of $r$ in it, we find that $\sigma_0 \psi_1=0$, which is a contradiction. This means that the system is inconsistent and admits no solutions. The source of this inconsistency is that $W_r$, $W_\phi$ and $W_Z$ can be determined in a unique manner from the algebraic equation~\eref{matrixform2}. They must however also be first derivatives of a smooth function $W(r,\phi,Z)$ and hence satisfy the Clairaut theorem on mixed derivatives. This contradicts~(\ref{reducedAa}-\ref{reducedAb}).

\section{Solutions of determining equations for Case 2: $\mathrm{rank}(M) = 2$}\label{SdeM2}

There are two main subcases to consider here: a) $\psi'(\phi)=0$, and b) $\mu(Z)=0$ while $\psi'(\phi) \neq 0$, so that we ensure that the determinant~\eref{detM} vanishes and thus the rank of $M$ is at most $2$. 

\subsection{Case 2a: $\psi'(\phi) = 0$}

It is understood again that $\psi(\phi)$ is set to zero. There are several ways for the rank to be equal to $2$. We recall the reduced row echelon form of the matrix $M$
\begin{equation}\label{Mreduced2a}
\left(\begin{array}{ccc}
0 & r^2 \mu(Z) & -\tau(\phi) \\
0 & \rho(r) & 0 \\
0 & 0 & \sigma(r) 
\end{array}\right)
\end{equation} 
The rank of a matrix is the largest size of its invertible square submatrices. Thus for the rank of the matrix~\eref{Mreduced2a} to be $2$, at least one of the three minors involving the second and third column must be non-zero. The possibilities are as follows: 
\begin{enumerate}[label=\arabic*)]
\item $\tau(\phi) \rho(r) \neq 0$, and then $\mu(Z)$ and $\sigma(r)$ are arbitrary;
\item $\mu(Z) \sigma(r) \neq 0$, and then $\tau(\phi)$ and $\rho(r)$ are arbitrary;
\item $\rho(r) \sigma(r) \neq 0$, and then $\mu(Z)$ and $\tau(\phi)$ are arbitrary.
\end{enumerate}
Let us consider these cases one by one.

\begin{enumerate}[label=\arabic*)]
\item  $\tau(\phi) \rho(r) \neq 0$.

From~\eref{reducedAa}, we have that $\tau(\phi)=\tau_0$ is a non-zero constant. Now recall that $\rho(r) \neq 0$ implies that $W_\phi = -\frac{1}{4} \mu'(Z) \tau(\phi)$, then notice that $\rho(r)W_\phi=0$. So $W_\phi=0$, and $\mu(Z)=\mu_0$ is a constant. It follows that $W_Z=0$. All of~\eref{reducedB} is then satisfied trivially. The solution for the magnetic field and the potential reads
\begin{equation}\label{rank2psi0cyl}
\fl W = W(r), \, B^r =0 , \, B^\phi =\frac{\tau_0}{r^3} + \frac{1}{2} \sigma'(r) , \, B^Z = \mu_0 r - \frac{1}{2}\rho'(r) .
\end{equation}
The integrals~\eref{cylintegrals} are determined by
\begin{eqnarray}
& s_1^r = 0, \quad s_1^\phi = \rho(r)-r^2 \mu_0 ,\quad s_1^Z = \tau_0, \nonumber \\
& m_1 =  \frac{1}{2} \left( \tau_0 \sigma(r) - r^2 \mu_0 \rho(r) -\left(\frac{\tau_0}{r}\right)^2 \right)+\frac{1}{4} \left( \rho(r)^2 + \mu_0^2 r^4 \right), \nonumber  \\
& s_2^r = 0, \quad s_2^\phi = \mu_0, \quad s_2^Z = \sigma(r)-\frac{\tau_0}{r^2}, \\
& m_2 = \frac{1}{2} \left( \rho(r) \mu_0 - \mu_0^2 r^2 -\frac{\tau_0 \sigma(r)}{r^2} \right) +\frac{1}{4} \left( \sigma(r)^2+\left(\frac{\tau_0}{r^2}\right)^2\right).\nonumber 
\end{eqnarray}

Recalling~\eref{transformB}, we express the system~\eref{rank2psi0cyl} in cartesian coordinates
\begin{eqnarray}
W &= W\left(\sqrt{x^2+y^2}\right), \nonumber \\
B^x &= -y \left( \frac{\tau_0}{\left( x^2+y^2 \right)^2} + S\left(\sqrt{x^2+y^2}\right) \right), \nonumber \\
B^y &= x \left( \frac{\tau_0}{\left( x^2+y^2 \right)^2} + S\left(\sqrt{x^2+y^2}\right) \right), \label{rank2psi0cart}\\
B^z &= \mu_0 - P\left(\sqrt{x^2+y^2}\right) \nonumber,
\end{eqnarray}
where $S(r) = \frac{\sigma'(r)}{2r}$ and $P(r) = \frac{\rho'(r)}{2r}$.

\item $\mu(Z) \sigma(r) \neq 0$.

The computation is very similar to the previous subcase. This time from~\eref{reducedAb} we see that $\mu'(Z)\sigma(r)=0$, so $\mu(Z)=\mu_0$ is a non-zero constant. Now recall that $\sigma(r) \neq 0$ implies that $W_Z = \frac{1}{4r^2} \tau'(\phi) \mu(Z)$, and notice that $\sigma(r)W_Z=0$. So $W_Z=0$, and $\tau(\phi)=\tau_0$ is a constant. It follows that $W_\phi=0$, and we have the same solution for $W(r,\phi,Z)$. The magnetic field is also the same, except that now $\rho(r)$ is arbitrary and $\sigma(r)$ is arbitrary and non-zero.

\item $\rho(r) \sigma(r) \neq 0$.

Recall that this directly implies that $\mu(Z)=\mu_0$ and $\tau(\phi)=\tau_0$ are constants. Once again~\eref{reducedB} is the same and the solutions are identical, except that neither $\rho(r)$ nor $\sigma(r)$ can be equal to zero.
\end{enumerate}
Thus the results for the case $\mathrm{rank } M=2$ and $\psi'(\phi) = 0$ take the form~\eref{rank2psi0cyl} (or, equivalently,~\eref{rank2psi0cart}). We notice that for the system~\eref{rank2psi0cyl} the two quadratic  integrals~\eref{cylintegrals} can be reduced to the first order integrals 
\begin{equation}\label{rank2psi0cyl1int}
\tilde{X}_1= 
p_\phi^A+\frac{\rho(r)}{2}-\frac{\mu_0 r^2}{2}, \quad \tilde{X}_2 = p_Z^A+\frac{\sigma(r)}{2}-\frac{\tau_0}{2 r^2}.
\end{equation}
Thus the system~\eref{rank2psi0cart} was already encountered in~\cite{Marchesiello_2015}, cf. equation~(76) therein. We notice that without any loss of generality we can absorb the constants $\tau_0$ and $\mu_0$ into a redefinition of $\sigma(r)$ and $\rho(r)$, i.e. set $\tau_0=\mu_0=0$ in~\eref{rank2psi0cyl}, \eref{rank2psi0cart} and~\eref{rank2psi0cyl1int}.

\subsection{Case 2b: $\mu(Z) = 0$, $\psi'(\phi) \neq 0$}

Under these assumptions equations~\eref{reducedB} directly imply that the variable $Z$ can be separated from the other two variables $r$ and $\phi$  in the potential $W$, i.e. 
\begin{equation}\label{W123}
W(r,\phi,Z) = W_{12}(r,\phi) + W_3(Z).
\end{equation}
The reduced row echelon form of $M$ becomes
\begin{equation}\label{redMcase2b}
\left(\begin{array}{ccc}
r \psi'(\phi) & r \rho(r) - \psi(\phi) & 0 \\
0 & 0 & \sigma(r) \\
0 & 0 & \tau(\phi)
\end{array}\right)
\end{equation}
Our assumption $\psi'(\phi) \neq 0$ implies that $r \rho(r) - \psi(\phi) \neq 0$. Thus to have $\mathrm{rank}\, M=2$ we have two possibilities, namely $\sigma(r) \neq 0$ or $\tau(\phi) \neq 0$. Either of them implies
\begin{equation}
W_Z  =0. \label{vanW3}
\end{equation}
Since the separation of the potential~\eref{W123} is defined up to an additive constant, we can set $W_3(Z)=0$, i.e. we have $W(r,\phi,Z)=W_{12}(r,\phi)\equiv W(r,\phi)$.

\begin{enumerate}[label=\arabic*)]
\item $\sigma(r) \neq 0$.

We first rewrite~\eref{reducedAa} in the following way:
\begin{equation}
r^3 \sigma'(r) + 2 \tau(\phi) - \frac{\tau'(\phi)}{\psi'(\phi)}(r \rho(r) - \psi(\phi)) = 0.
\end{equation}
Differentiation with respect to $\phi$ leads to the equation:
\begin{equation} \label{separablemess}
\fl 3 \tau'(\phi) + \psi(\phi) \frac{\tau''(\phi) \psi'(\phi) - \tau'(\phi) \psi''(\phi)}{\psi'(\phi)^2} = r \rho(r) \frac{\tau''(\phi) \psi'(\phi) - \tau'(\phi) \psi''(\phi)}{\psi'(\phi)^2}.
\end{equation}
If $\tau''(\phi) \psi'(\phi) - \tau'(\phi) \psi''(\phi) \neq 0$ we can separate the variables $r$ and $\phi$. If instead this expression vanishes, we directly conclude from~\eref{separablemess} that $\tau'(\phi)=0$, thus $\tau(\phi)=\tau_0$ is a constant. We study both situations separately.

\begin{enumerate}[label=1.\arabic*)]
\item $\tau''(\phi) \psi'(\phi) - \tau'(\phi) \psi''(\phi) = 0$, i.e. $\tau(\phi)=\tau_0$.

Equation~\eref{reducedAa} now reads $r^3 \sigma'(r) = -2 \tau_0$; thus, we have $\sigma(r) = \frac{\tau_0}{r^2} + \sigma_0$. This reduces the system~(\ref{reducedAa}--\ref{matrixform2}) to the following two equations
\begin{eqnarray}
r\psi'(\phi)W_r + \left( r \rho(r) - \psi(\phi) \right)W_\phi = 0,  \label{mcsween} \\
\psi'(\phi) \left( -3 \psi''(\phi) + r^3 \rho''(r) - r^2 \rho'(r) + r \rho(r) - 4 \psi(\phi) \right)  \nonumber \\
+ \psi'''(\phi) \left( r \rho(r) - \psi(\phi) \right) - 4 r^5 W_{r \phi} - 8 r^4 W_\phi  = 0.
\end{eqnarray}
 The magnetic field takes the form
\begin{equation}\label{r2mu0B}
B^\phi = 0, \quad B^r = 0, \quad  B^Z = -\rho'(r)-\frac{\psi''(\phi)+\psi(\phi)}{2 r^2}.
\end{equation}
Thus the motion of the system splits into a motion in the $xy$-plane under the influence of the potential $W(r,\phi)$ and the perpendicular magnetic field $B^Z(r,\phi)$  (a problem discussed by McSween and Winternitz in polar coordinates in~\cite{mcsween_2000}) plus a free motion in the $z$-direction. The integral $X_2$ reduces to a first order one
\begin{equation}
\tilde{X}_2=p_Z^A+\frac{\sigma_0}{2}
\end{equation}
and in a suitably chosen gauge becomes simply $p_Z$.

\item $\tau''(\phi) \psi'(\phi) - \tau'(\phi) \psi''(\phi) \neq 0$.

In this case we can separate the variables $r$ and $\phi$ in~\eref{separablemess}, arriving at the equations 
\begin{equation}
\frac{3 \tau'(\phi) \psi'(\phi)^2}{\tau''(\phi) \psi'(\phi) - \tau'(\phi) \psi''(\phi)} + \psi(\phi) = \rho_0 = r \rho(r),
\end{equation}
where $\rho_0$ is the separation constant. Solving them we find 
\begin{equation}
\rho(r)=\frac{\rho_0}{r},\qquad\tau(\phi) = \tau_0 + \frac{\tau_1}{(\psi(\phi)-\rho_0)^2}.
\end{equation} 
From equation~\eref{reducedAa} we find $\sigma(r) = \frac{\tau_0}{r^2} + \sigma_0$.

Next, we insert these results into the remaining equations~(\ref{reducedB}-\ref{matrixform2}) and find two equations which read
\begin{eqnarray}\label{12remeqs}
\fl r \psi'(\phi) W_r + \left( \rho_0 - \psi(\phi) \right)W_\phi &= 0, \nonumber \\
\fl -3 \psi'(\phi) \psi''(\phi) - 4 \psi'(\phi) \left( \psi(\phi) - \rho_0 \right) - \psi'''(\psi(\phi) - \rho_0)  \\
\fl - \frac{4 \tau_1^2}{\left( \psi(\phi) - \rho_0 \right)^5} \psi'(\phi) - 4r^5 W_{r \phi} - 8r^4 W_\phi &= 0.\nonumber
\end{eqnarray}
We can rewrite $\beta(\phi) = \psi(\phi) - \rho_0$ and integrate the second equation once with respect to $\phi$, arriving at the system
\numparts \label{betabeta}
\begin{eqnarray}
r \beta'(\phi) W_r - \beta(\phi)W_\phi & = 0 \label{betabeta1} \\
-\beta(\phi) \beta''(\phi) - \beta'(\phi)^2 - 2 \beta(\phi)^2 + \frac{\tau_1^2}{\beta(\phi)^4} & \nonumber \\ 
- 4r^5 W_r - 8r^4 W(r,\phi) - f(r) & = 0. \label{betabeta2}
\end{eqnarray}
\endnumparts
Substituting for $W_r$ from~\eref{betabeta1} into~\eref{betabeta2} we find expressions for both $W_r$ and $W_\phi$. Substituting them into~\eref{12remeqs} we find that 
\begin{equation}
f(r) = \frac{f_1}{4}+f_2 r^4
\end{equation}
in~\eref{betabeta2}, where $f_1$, $f_2$ are integration constants. Next, we find solving~\eref{betabeta1} the explicit form of the potential in terms of the yet unknown function $\beta(\phi)$
\begin{equation}\label{12Wsolved}
\fl W = -\frac{f_2}{8}+\frac{\tilde{W}(\phi)}{r^2}+\frac{\beta(\phi) \beta''(\phi)+\beta'(\phi)^2+\frac{f_1}{4}-\frac{\tau_1^2}{\beta(\phi)^4}+2 \beta(\phi)^2}{ 8 r^4}.
\end{equation}
The function $\tilde{W}(\phi)$ is determined by~\eref{betabeta2} and up to a constant shift of the potential reads
\begin{equation}
\tilde{W}(\phi) = \frac{W_0}{\beta(\phi)^2}+\frac{f_2}{8},
\end{equation}
where $W_0$ is an arbitrary constant. The potential thus becomes fully determined, 
\begin{equation}\label{12Wsolved2}
\fl W = \frac{W_0}{r^2 \beta(\phi)^2}+\frac{\beta(\phi) \beta''(\phi)+\beta'(\phi)^2+\frac{f_1}{4}-\frac{\tau_1^2}{\beta(\phi)^4}+2 \beta(\phi)^2}{ 8 r^4}.
\end{equation}
The sole remaining equation~\eref{betabeta2} becomes an equation for the uknown function $\beta(\phi)$ only, namely
\begin{eqnarray} \label{betasimple}
\fl \beta'(\phi) \left( 7 \beta(\phi)\beta''(\phi)+4\beta'(\phi)^2 + 12 \beta(\phi)^2 + f_1 \right) + \beta(\phi)^2\beta'''(\phi)=0.
\end{eqnarray}
This equation can be integrated twice, i.e. reduced to a first order ODE. In order to do this we must multiply by $\beta(\phi)$ and integrate, then multiply by $\beta'(\phi)\beta(\phi)$ and integrate again. The result is 
\begin{equation} \label{betafirstorder}
4 \beta(\phi)^4  \beta'(\phi)^2 + 4 \beta(\phi)^6 - 4 \beta_1 \beta(\phi)^2+ f_1 \beta(\phi)^4 = \beta_2
\end{equation}
where $\beta_1,\beta_2$ are the constants of integration. Substituting $\gamma(\phi)=\beta(\phi)^2$ we can re-express it as
\begin{equation} \label{gammaeq}
\gamma(\phi) \gamma'(\phi)^2+4 \gamma(\phi)^3-4 \beta_1 \gamma(\phi)+f_1 \gamma(\phi)^2 = \beta_2.
\end{equation}
In the special case where $\beta_2=0$, it is possible to solve this equation and the solution is 
\begin{equation}\label{gammaeqsolspec}
\gamma(\phi) = \frac{\sqrt{ 64 \beta_1+f_1^2}\sin\left( 2 (\phi-\phi_0) \right)-f_1}{8}.
\end{equation}
Under the assumption that $f_1<0$, $\frac{f_1}{8}<\beta_1<0$ the function $\beta(\phi)$ is well defined, bounded and positive,
\begin{equation}
\beta(\phi)=\sqrt{\frac{\sqrt{ 64 \beta_1+f_1^2}\sin\left( 2 (\phi-\phi_0) \right)-f_1}{8}}.
\end{equation}

To our knowledge for $\beta_2 \neq 0$ the solution of~\eref{gammaeq} cannot be expressed in terms of known analytic functions.

The magnetic field is also expressed in terms of the function $\beta(\phi)$ and reads
\begin{eqnarray}\label{12B}
B^r & = -\tau_1 \frac{\sqrt{4 \beta_1 \beta(\phi)^2+\beta_2-4\beta(\phi)^6- f_1 \beta(\phi)^4}}{2 r^2 \beta(\phi)^5}, \nonumber \\
B^\phi & = \frac{\tau_1}{r^3 \beta(\phi)^2}, \quad B^Z = \frac{2 \beta_1 \beta(\phi)^2+\beta_2}{4 r^2 \beta(\phi)^5}. 
\end{eqnarray}
(The sign of the square root depends on the choice of the branch of the square root of $\beta'(\phi)$ in~\eref{betafirstorder}.)

For example, for the solution~\eref{gammaeqsolspec} we find the following structure of the magnetic field
\begin{eqnarray}
B^r & = -\frac{8 \tau_1 \sqrt{f_1^2+64 \beta_1} \cos\left(2 (\phi- \phi_0)\right)}{r^2 \left(\sqrt{f_1^2+64 \beta_1} \sin\left(2 (\phi- \phi_0)\right)-f_1\right)^2}, \nonumber \\
B^\phi & = \frac{8 \tau_1}{r^3  \sqrt{ 64 \beta_1+f_1^2}\sin\left( 2 (\phi-\phi_0) \right)-f_1 },\\ \nonumber
B^Z & = \frac{\beta_1}{2 r^2} \left(\frac{\sqrt{ 64 \beta_1+f_1^2}\sin\left( 2 (\phi-\phi_0) \right)-f_1}{8}\right)^{-\frac{3}{2}}
\end{eqnarray}

Using~\eref{betafirstorder} the potential~\eref{12Wsolved2} simplifies to an explicit function of $\beta(\phi)$,
\begin{equation}\label{12W}
W = \frac{W_0}{r^2 \beta(\phi)^2}-\frac{4 \tau_1^2+\beta_2}{32 \beta(\phi)^4 r^4}.
\end{equation}
In particular, for the solution~\eref{gammaeqsolspec} we find
\begin{eqnarray}
W = & \frac{8 W_0}{r^2 \left(\sqrt{f_1^2+64 \beta_1} \sin\left(2(\phi-\phi_0)\right)-f_1\right)} \nonumber \\ 
 & -\frac{8 \tau_1^2}{r^4 \left(\sqrt{f_1^2+64 \beta_1} \sin\left(2(\phi-\phi_0)\right)-f_1\right)^2}.
\end{eqnarray}

The integrals~\eref{cylintegrals} are determined by
\begin{eqnarray}
s_1^r & = \frac{\sqrt{4 \beta_1 \beta(\phi)^2+\beta_2-4 \beta(\phi)^6-\beta(\phi)^4 f_1}}{2 \beta(\phi)^2}, \nonumber \\
s_1^\phi & = -\frac{\beta(\phi)}{r}, \qquad s_1^Z = \frac{\beta(\phi)^2 \tau_0+\tau_1}{\beta(\phi)^2}, \nonumber \\
m_1 & = \frac{2 W_0}{\left(\beta(\phi)\right)^2}-\frac{4 \beta(\phi)^2 \tau_0 \tau_1+2 \beta_1 \beta(\phi)^2+4 \tau_1^2+\beta_2}{8 \beta(\phi)^4 r^2},  
\\ \nonumber 
s_2^r & = 0 ,\qquad  s_2^\phi = 0, \qquad s_2^Z  = \sigma_0-\frac{\tau_1}{r^2 \beta(\phi)^2}, \\ \nonumber 
m_2 & = \frac{\tau_1}{\beta(\phi)^2 r^2} \left( \frac{\tau_1}{4 \beta(\phi)^2 r^2} - \frac{\sigma_0}{2} \right).
\end{eqnarray}
From the presence of the parameters $\tau_0$ and $\sigma_0$ in the integrals on which the magnetic field and the potential do not depend we deduce that the integral $X_2$ of the system~(\ref{12B}--\ref{12W}) actually can be reduced to a first order one, namely
\begin{equation}
\tilde{X}_2= p_Z^A-\frac{\tau_1}{2 \beta(\phi)^2 r^2}.
\end{equation}
\end{enumerate}

\item $\tau(\phi) \neq 0$.

In this case it is now understood that there is no constraint on $\sigma(r)$ yet. But in the previous case we never actually considered a case where $\tau(\phi)$ would vanish, and there was no division by $\sigma(r)$, so we can follow the same splitting as well as some of the same results. So the first subcase is once again the polar case treated in Ref. \cite{mcsween_2000} but with $\tau(\phi) \neq 0$, and the second subcase is again the same as in~\eref{12B} and~\eref{12W} while taking~(\ref{betabeta1}--\ref{betabeta2}) into account.
\end{enumerate}

\section{Solutions of determining equations for Case 3: $\mathrm{rank}(M) = 1$}\label{SdeM1}

Once again there are only two consistent ways for the determinant of $M$ to vanish, i.e. $\psi'(\phi) = 0$ which without loss of generality becomes $\psi(\phi)=0$, and $\mu(Z) = 0$ while $\psi'(\phi) \neq 0$.

\subsection{Case 3a: $\psi'(\phi) = 0$}

We have the same reduced row echelon form~\eref{Mreduced2a} for $M$. This time around we ask the rank to be $1$, so every minor of size $2$ has to vanish, but there has to remain at least one non-zero entry. There are four possibilities, one for each function to individually be non-zero,
\begin{enumerate}[label=\arabic*)]
\item $\mu(Z) \neq 0$, this implies that $\sigma(r)=0$ and $\rho(r) \tau(\phi)=0$,
\item $\mu(Z)=0$, $\tau(\phi) \neq 0$, this implies that $\rho(r)=0$,
\item $\mu(Z)=0$, $\tau(\phi) = 0$ and $\rho(r)\neq 0$, this implies that $\sigma(r)=0$,
\item $\mu(Z)=0$, $\tau(\phi) = 0$ and $\rho(r)=0$, this implies that $\sigma(r)\neq 0$.
\end{enumerate}
Let us now consider these cases separately
\begin{enumerate}[label=\arabic*)]
\item $\mu(Z) \neq 0$, $\sigma(r)=0$, $\rho(r) \tau(\phi) = 0$.

We use the fact that $\rho(r) W_\phi=0$, which further splits the problem into two subcases.
\begin{enumerate}
\item Let's first consider what happens when $\rho(r)=0$. Plugging everything we know into~~(\ref{reducedAa}-\ref{reducedAb}),~\eref{reducedB} and~\eref{matrixform2}, we have the remaining four equations:
\numparts\label{taumusys}
\begin{eqnarray} 
W_{r \phi} &= -\frac{2}{r}W_\phi + \frac{1}{2r^5} \tau'(\phi) \tau(\phi) - \frac{1}{2r} \tau(\phi) \mu'(Z) , \label{appeq1}\\
W_{\phi Z} &= -\frac{1}{4} \mu''(Z) \tau(\phi) - \frac{1}{4r^2} \tau''(\phi) \mu(Z), \label{appeq2} \\
W_{r Z} &=  - \frac{r}{2} \mu'(Z) \mu(Z) + \frac{1}{2r^3} \mu(Z) \tau'(\phi), \label{appeq3} \\
r^2 \mu(&Z)W_\phi - \tau(\phi)W_Z = 0. \label{appeq4}
\end{eqnarray}
\endnumparts
We introduce $M'(Z)=\mu(Z)$ and $T'(\phi)=\tau(\phi)$. Integrating~\eref{appeq2} with respect to $Z$ and $\phi$ we find an expression for the potential in terms of two functions of two variables each:
\begin{equation}
\fl W(r,\phi,Z) = -\frac{1}{4r^2}\tau'(\phi)M(Z) - \frac{1}{4}T(\phi)\mu'(Z) + F_1(r,\phi) + F_2(r,Z).
\end{equation}
This expression for $W$ we substitute into~\eref{appeq3}, finding $F_2(r,Z)$. Inserting it into~\eref{appeq1} we find $F_1(r,\phi)$. Thus we arrive at the explicit form of the potential
\begin{eqnarray}\label{potentialbigtaumu}
\fl W(r,\phi,Z) &= -\frac{1}{4r^2}T''(\phi)M(Z) - \frac{1}{4}T(\phi) M''(Z) - \frac{r^2}{8}M'(Z)^2 \nonumber \\
\fl &- \frac{1}{8r^4}T'(\phi)^2 + W_1(r) + \frac{1}{r^2}W_2(\phi) + W_3(Z),
\end{eqnarray}
We are left with a single equation~\eref{appeq4} to solve, which simplifies to

\begin{equation} \label{bigtaumu}
\fl T(\phi)T'(\phi)M'''(Z) - M(Z)M'(Z)T'''(\phi) = 4 \left(T'(\phi)W_3'(Z) -M'(Z)W_2'(\phi) \right).
\end{equation}\smallskip
% the sign was flipped, in agreement with my Maple calculation

If we assume that $\tau(\phi) \neq 0$, it is possible to separate the variables $\phi$ and $Z$ by dividing the above expression by $M'(Z)T'(\phi)$ and then differentiating with respect to $\phi$ and $Z$. This leads to the following condition:
\begin{equation}
\fl \frac{T''''(\phi)T'(\phi) - T'''(\phi)T''(\phi)}{T'(\phi)^3} = - 3 C = \frac{M''''(Z)M'(Z)-M'''(Z)M''(Z)}{M'(Z)^3},
\end{equation}
for some separation constant $C$. Reducing the order of the separated equations, we find that
\begin{eqnarray}\label{WeiEqs}
M'(Z)^2 & = C M(Z)^3+C_1 M(Z)^2+C_2 M(Z)+C_3, \\ \nonumber  T'(\phi)^2 & = C T(\phi)^3+\tilde{C}_1 T(\phi)^2+\tilde{C}_2 T(\phi)+\tilde{C}_3
\end{eqnarray}
where $C_1,C_2,C_3,\tilde{C}_1,\tilde{C}_2,\tilde{C}_3$ are constants of integration. For $C\neq 0$ the right hand side of~\eref{WeiEqs} can be rewritten in terms of the roots of third order polynomials, e.g.
\numparts
\begin{eqnarray}\label{WeiEqsRootsM}
M'(Z)^2 & = C (M-M_1)(M-M_2)(M-M_3), \\   T'(\phi)^2 & = C (T-T_1)(T-T_2)(T-T_3).\label{WeiEqsRootsT}
\end{eqnarray}
\endnumparts
We are interested in real solutions. If the constants in~\eref{WeiEqs} are real then $M_i$  are either also real or one of them is real and the other two complex and mutually complex conjugate, and similarly for $M_i$, e.g.
\begin{eqnarray}
\fl M_1\in \R, \quad M_2=p+\mathrm{i} q, \quad M_3=p-\mathrm{i} q,\quad p,q\in\R, \, q>0.
\end{eqnarray}
Let us focus on~\eref{WeiEqsRootsM}. If the three roots are all different and real then the solution of~\eref{WeiEqsRootsM} is best expressed in terms of ${\mathrm{sn}}^2\,(u,k)$ where ${\mathrm{sn}}\,(u,k)$ is the Jacobi elliptic sine function. The argument $u$ is proportional to $Z$ in~\eref{WeiEqsRootsM}, the modulus $k$ determines the real and imaginary periods of the Jacobi elliptic function. The values of $k$ can be restricted to $0<k<1$. The solutions of~\eref{WeiEqsRootsM} will be real in regions where the r.h.s is nonnegative. If two of the roots $M_i$ are complex the solutions $M(Z)$ are best expressed in terms of the elliptic cosine function ${\mathrm cn}\,(u,k)$. In the case of double roots the elliptic functions reduce to elementary ones and the module $k$ takes one of the limiting values $k=0$ or $k=1$.

We shall not go into further details here and simply refer to the book~\cite{ByrdFriedman} for a comprehensive and detailed review. Here we limit ourselves to several examples.\medskip

\begin{description}
\item{{\bf Example 1.}} $M_1>M_2>M(Z)>M_3$, $C>0$.

We put 
\begin{equation}\label{WeiEqsRootsMsol1}
\fl M(Z)=(M_2-M_3) {\mathrm{sn}}^2(u,k)+M_3, \; k^2=\frac{M_2-M_3}{M_1-M_3}, \; u=\frac{\sqrt{C(M_1-M_3)}}{2} Z.
\end{equation} Equation~\eref{WeiEqsRootsM} for $M(Z)$ then reduces to the  first order ODE defining the Jacobi sine function,
\begin{equation}\label{WeiEqsRootsMs}
\fl \left( \frac{\mathrm{d}\, {\mathrm{sn}}\,(u,k)}{\mathrm{d}\, u}\right)^2 = \left(1-{\mathrm{sn}}^2\,(u,k)\right)\left(1-k^2 {\mathrm{sn}}^2\,(u,k)\right) .
\end{equation}
Notice that we have $0<k^2<1$ and the solution is real and finite, satisfying $M_3\leq M(Z)\leq M_2$ (since we have $0\leq {\mathrm{sn}}^2\,(u,k) \leq 1$).\medskip

\item{{\bf Example 2.}} $M(Z)>M_1>M_2>M_3$, $C>0$.

We put 
\begin{equation}\label{WeiEqsRootsMsol2}
\fl M(Z)=\frac{M_1-M_2 \, {\mathrm{sn}}^2(u,k)}{1-{\mathrm{sn}}^2(u,k)},\, k^2=\frac{M_2-M_3}{M_1-M_3}, \, u=\frac{\sqrt{C(M_2-M_3)}}{2} Z,
\end{equation}
and this reduces~\eref{WeiEqsRootsM} to~\eref{WeiEqsRootsMs}. We again have $0<k^2<1$. The solution~\eref{WeiEqsRootsMsol2} is real, periodic and singular with simple poles given by ${\mathrm{sn}}^2(u,k)=1$. \medskip
\end{description}

The real period of all Jacobi functions depends on a number $K(k)$. In particular we have 
\begin{equation}
\fl {\mathrm{sn}}\,(u,k)={\mathrm{sn}}\,(u+4K,k), \quad K=\frac{\pi}{2} F\left(\frac{1}{2},\frac{1}{2},1; k^2\right)
\end{equation}
where $F\left(a,b,c;z\right)$ is the Gauss hypergeometric function.

The next two examples are elementary solutions, i.e. cases when two roots of the polynomial in~\eref{WeiEqsRootsM} coincide. They can either be obtained by direct integration of~\eref{WeiEqsRootsM} or as special limiting cases of solutions~\eref{WeiEqsRootsMsol1} and~\eref{WeiEqsRootsMsol2}.\medskip

\begin{description}
\item{{\bf Example 3.}} $M_1=M_2>M(Z)>M_3$, $C>0$

Putting $M_1=M_2$ in~\eref{WeiEqsRootsM} and using ${\mathrm{sn}}\,(u,1)=\tanh u$ we obtain
\begin{equation}\label{WeiEqsRootsMsol3}
\fl M(Z)=(M_1-M_3) \tanh^2(u)+M_3, \quad u=\frac{\sqrt{C(M_1-M_3)} Z}{2}.
\end{equation}
The solution~\eref{WeiEqsRootsMsol3} is real, has no singularities on the real axis and satisfies $M(0)=M_3$, $\lim_{u\rightarrow \pm \infty} M(u)=M_1$. \medskip

\item{{\bf Example 4.}} $M(Z)>M_1>M_2=M_3$, $C>0$

For $M_2=M_3$ we have $k^2=0$ in~\eref{WeiEqsRootsMsol2} and use ${\mathrm{sn}}\,(u,0)=\sin u$. The solution~\eref{WeiEqsRootsMsol2} reduces to
\begin{equation}\label{WeiEqsRootsMsol4}
M(Z)=\frac{M_1-M_2 \sin^2 u}{1-\sin^2 u} , \quad u=\frac{\sqrt{C(M_1-M_2)} }{2} Z.
\end{equation}
The solution~\eref{WeiEqsRootsMsol4} is real, periodic and has simple poles at $\sin^2u=1$, i.e. $u=\left(j+\frac{1}{2}\right) \pi$, $j\in\mathbb{Z}$.\medskip
\end{description}

Alternatively, we can transform equation~\eref{WeiEqs} for $C\neq 0$ into the Weierstrass form by putting
\begin{equation}
M(Z)=\frac{4}{C} S(Z)-\frac{2\tilde{C_1}}{3C}, \quad S(Z)=\wp\left( 2Z,g_2,g_3\right)
\end{equation}
where $\wp\left( 2Z,g_2,g_3\right)$ is the Weierstrass elliptic function and $g_2$, $g_3$ are expressed in terms of the arbitrary constants $\tilde{C}_1$, $\tilde{C}_2$, $\tilde{C}_3$ and $C$. Similar transformations can be performed for the function $T(\phi)$ to solve the equation~\eref{WeiEqs} for $T(\phi)$ in terms of Jacobi or Weierstrass elliptic functions or their degenerate cases.

The functions $W_2(\phi),W_3(Z)$ are determined by equation~\eref{bigtaumu} which using~\eref{WeiEqs} can be rewritten in a separated form
\begin{equation}\label{bigtaumuA}
\frac{W_2'(\phi)}{T'(\phi)}+\frac{C_1}{4} T(\phi)=\frac{W_3'(Z)}{M'(Z)}+\frac{\tilde{C}_1}{4} M(Z) = w_0.
\end{equation}
Thus the potential~\eref{potentialbigtaumu} is determined by solutions of~\eref{bigtaumuA}, 
\begin{eqnarray}
W_2(\phi) & = -\frac{C_1}{8} \left(T(\phi)\right)^2+w_0 T(\phi), \\ \nonumber 
W_3(Z) & = -\frac{\tilde{C}_1}{8} \left(M(Z)\right)^2+w_0 M(Z),
\end{eqnarray}
where the integration constants were without loss of generality absorbed into the function $W_1(r)$.

The magnetic field and the functions determining the integrals are expressed in terms of solutions $M(Z)$, $T(\phi)$, $W_2(\phi)$, $W_3(Z)$ of equation~\eref{bigtaumu} as follows
\begin{eqnarray}\label{weierstrassBsm}
\fl B^r & = \frac{T''(\phi)}{2 r^2} - \frac{r^2 M''(Z)}{2}, \qquad B^\phi = \frac{T'(\phi)}{r^3}, \qquad B^Z =  r M'(Z), \nonumber  \\
\fl s_1^r & = 0, \qquad s_1^\phi = -r^2 M'(Z),\qquad  s_1^Z = T'(\phi), \nonumber \\ 
\fl s_2^r & = 0, \qquad s_2^\phi = M'(Z), \qquad s_2^Z = -\frac{T'(\phi)}{r^2},\\ \nonumber 
\fl m_1 & = \frac{r^4}{4} \left(M'(Z)\right)^2-\frac{\left(T'(\phi)\right)^2}{2 r^2}-\frac{M(Z) T''(\phi)}{2} +2 W_2(\phi), \\ \nonumber
\fl m_2 & = -\frac{r^2 \left(M'(Z)\right)^2}{2} +\frac{\left(T'(\phi)\right)^2}{4 r^4}-\frac{M''(Z) T(\phi)}{2} +2 W_3(Z).
\end{eqnarray}

If $C=0$, we find solutions of~\eref{WeiEqs} expressed in terms of exponentials and trigonometric functions. Choosing e.g.\footnote{By a different choice of the integration constants we can write $\mu$ periodic in $Z$, i.e. linear combination of sine and cosine. The choice of trigonometric vs. exponential functions is governed by the sign of the constants $C_1$ and $\tilde{C}_1$.} the periodic solution for $\tau$ and unbounded solution for $\mu$
\begin{eqnarray}
\fl \mu(Z) & = k_1 {\mathrm{e}}^{k_0 Z} + k_2 {\mathrm{e}}^{-k_0 Z}, \qquad \tau(\phi) &= \tilde{k}_1 \cos(\tilde{k}_0 \phi) + \tilde{k}_2 \sin(\tilde{k}_0 \phi), 
\end{eqnarray}
the corresponding functions $T(\phi)$ and $M(Z)$ read
\begin{eqnarray}\label{C0TM}
M(Z) & = \frac{k_1 {\mathrm{e}}^{k_0 Z}-k_2 {\mathrm{e}}^{-k_0 Z}+k_3}{k_0}, \qquad C_1 = k_0^2, \\
T(\phi) & = \frac{\tilde{k}_1 \sin(\tilde{k}_0 \phi)-\tilde{k}_2 \cos(\tilde{k}_0\phi)+\tilde{k}_3}{\tilde{k}_0}, \quad  \tilde{C}_1 = -\tilde{k}_0^2 \nonumber
\end{eqnarray}
where $k_0$, $k_1$, $k_2$, $k_3$, $\tilde{k}_0$, $\tilde{k}_1$, $\tilde{k}_2$ and $\tilde{k}_3$  are arbitrary parameters (replacing the arbitrary integration constants $C_j,\tilde{C}_j$ of~\eref{WeiEqs}). The magnetic field and the potential take the form
\begin{eqnarray}\label{p0C0B}
\fl B^r & = \tilde{k}_0 \frac{\tilde{k}_2 \cos(\tilde{k}_0 \phi)-\tilde{k}_1 \sin(\tilde{k}_0 \phi)}{2 r^2}+k_0 \frac{\left(k_2 {\mathrm{e}}^{-k_0 Z}- k_1 {\mathrm{e}}^{k_0 Z}\right)r^2 }{2}, \nonumber \\
\fl B^\phi & = \frac{\tilde{k}_1 \cos(\tilde{k}_0 \phi)+\tilde{k}_2 \sin(\tilde{k}_0 \phi)}{r^3}, \quad 
B^Z = \left(k_1 {\mathrm{e}}^{k_0 Z}+k_2 {\mathrm{e}}^{-k_0 Z}\right) r, \nonumber \\
\fl W & = W_1(r)-\frac{r^2 (M'(Z))^2}{8}  +\frac{1}{8} \tilde{k}_0^2 M(Z)^2+w_0 M(Z)+ \\ \nonumber
\fl &+\frac{8 w_0 T(\phi)- k_0^2 T(\phi)^2-2 M(Z) T''(\phi)}{8 r^2}-\frac{(T'(\phi))^2}{8 r^4} -\frac{ M''(Z) T(\phi)}{4}
%\fl  W & = W_1(r)-\left( k_1 {\mathrm{e}}^{k_0 Z}+k_2 {\mathrm{e}}^{-k_0 Z}\right)^2 \frac{r^2}{8}+ \\ 
%\fl & \nonumber
%+(k_1^2 {\mathrm{e}}^{2 k_0 Z}+k_2^2 {\mathrm{e}}^{-2 k_0 Z}) \frac{\tilde{k}_0^2}{8 k_0^2}
%+(k_1 {\mathrm{e}}^{k_0 Z}-k_2 {\mathrm{e}}^{-k_0 Z}) \frac{w_1}{k_0} - \\ 
%\fl & \nonumber
%-\left(k_1 {\mathrm{e}}^{k_0 Z}-k_2 {\mathrm{e}}^{-k_0 Z}\right) \left(\tilde{k}_1 \sin(\tilde{k}_0 \phi)-\tilde{k}_2 \cos(\tilde{k}_0 \phi)+w_2\right) \frac{k_0}{4 \tilde{k}_0}+ \\ 
%\fl & \nonumber
%+\frac{1}{r^2} \left( \frac{k_0^2 w_2}{4 \tilde{k}_0^2} \left(\tilde{k}_2 \cos(\tilde{k}_0 \phi)-\tilde{k}_1 \sin(\tilde{k}_0 \phi)\right) +w_2+ \right. \\ 
%\fl \nonumber & \left. + \frac{w_1}{\tilde{k}_0} \left(\tilde{k}_1 \sin(\tilde{k}_0 \phi)-\tilde{k}_2 \cos(\tilde{k}_0 \phi)\right) +
%\frac{k_0^2}{8 \tilde{k}_0^2} \left(\tilde{k}_1 \cos(\tilde{k}_0 \phi)+\tilde{k}_2 \sin(\tilde{k}_0 \phi)\right)^2
%+ \right. \\ 
%\fl & \nonumber \left. 
% +\frac{\tilde{k}_0}{4 k_0} \left( k_1 {\mathrm{e}}^{k_0 Z} \left(\tilde{k}_1 \sin(\tilde{k}_0 \phi)-\tilde{k}_2 \cos(\tilde{k}_0 \phi)\right)+k_2  {\mathrm{e}}^{-k_0 Z} \left(\tilde{k}_1 \sin(\tilde{k}_0 \phi)+\tilde{k}_2 \cos(\tilde{k}_0 \phi)\right) \right) - \right. \\ 
% \fl & \nonumber \left. - \frac{k_0^2 \left( \tilde{k}_1^2 + \tilde{k}_2^2 \right) }{16 \tilde{k}_0^2} 
%\right)
%-\frac{1}{8 r^4}\left( \tilde{k}_1 \cos(\tilde{k}_0 \phi)+\tilde{k}_2 \sin(\tilde{k}_0 \phi)\right)^2.
\end{eqnarray} 
where substitution~\eref{C0TM} is assumed. The integrals of motion are determined by the functions~\eref{weierstrassBsm}, as above.

In the case $\tau(\phi)=0$ the solution is much more straightforward and we immediately arrive at the potential and the magnetic field 
\begin{eqnarray}\label{p0t0}
\nonumber W &= W_1(r) - \frac{r^2}{8}\mu(Z)^2 + W_3(Z),\\
B^r &= - \frac{r^2}{2}\mu'(Z) , \quad B^\phi =0 , \quad B^Z = r \mu(Z),
\end{eqnarray}
where $\mu(Z) \neq 0$ is an arbitrary  nonvanishing function.  The integrals are given by
\begin{eqnarray}
\fl s_1^r  & = 0, \quad s_1^\phi  = -r^2 \mu(z), \quad s_1^Z  = 0, \quad m_1  = \frac{r^4}{4} \left(\mu(z)\right)^2, \\ \nonumber 
\fl s_2^r  & = 0, \quad s_2^\phi  = \mu(z), \quad s_2^Z  = 0, \quad m_2  = -\frac{ r^2}{2} \left(\mu(z)\right)^2+2 W_3(Z).
\end{eqnarray}
Transforming the system into cartesian coordinates, we find
\begin{eqnarray}
W &= W_1 \left( \sqrt{x^2 + y^2} \right) - \frac{x^2+y^2}{8} \mu(z)^2 + W_3(z), \nonumber \\
\vec B &= \left( -\frac{x}{2}\mu'(z), -\frac{y}{2}\mu'(z), \mu(z)\right). 
\end{eqnarray}
Obviously, for this system the integral $X_1$ reduces to a first order one
\begin{equation}
\tilde{X}_1 = p_\phi^A-\frac{r^2}{2} \mu(z)
\end{equation}
since the magnetic field and the potential are invariant with respect to rotations around $z$-axis.

\item On the other hand if we have $\rho(r) \neq 0$, thus $W_\phi=0$,  $\tau(\phi)=0$ and there is only one remaining equation to be solved, namely
\begin{equation}\label{Wrztau0}
W_{r Z} = -\frac{r}{2}\mu'(Z)\mu(Z) + \frac{1}{4}\mu'(Z)\rho'(r).
\end{equation}
Solving for the potential, we conclude that both $\mu(Z)$ and $\rho(r)$ remain arbitrary nonvanishing functions, and the potential and magnetic field read
\begin{eqnarray}\label{Wrztau0sol}
\fl W &= W_1(r) - \frac{r^2}{8}\mu(Z)^2 + \frac{1}{4}\rho(r)\mu(Z) + W_3(Z),\\
\fl B^r &= - \frac{r^2}{2}\mu'(Z) , \qquad B^\phi =0 , \qquad B^Z = r \mu(Z) - \frac{1}{2}\rho'(r) . \nonumber
\end{eqnarray}
In cartesian coordinates they become
\begin{eqnarray}
\fl W = W_1 \left( \sqrt{x^2 + y^2} \right) - \frac{x^2+y^2}{8} \mu(z)^2 + \frac{1}{4}\rho \left( \sqrt{x^2+y^2} \right) \mu(z) + W_3(z), \nonumber \\
\fl \vec B = \left( -\frac{x}{2}\mu'(z), -\frac{y}{2}\mu'(z), \mu(z) - P \left( \sqrt{x^2+y^2} \right) \right), 
\end{eqnarray}
where $P(r)=\frac{\rho'(r)}{2r}$.  

The integrals are determined by
\begin{eqnarray}
\fl s_1^r & = 0, \, s_1^\phi  = \rho(r)-r^2 \mu(Z), \, s_1^Z  = 0,\nonumber \\
\fl  m_1 & = \frac{r^4}{4}  \mu(Z)^2-\frac{r^2}{2} \mu(Z) \rho(r) +\frac{\rho(r)^2}{4}, \\ \nonumber 
\fl s_2^r & = 0, \, s_2^\phi  = \mu(Z), \, s_2^Z  = 0,\, m_2  = -\frac{r^2}{2} \mu(Z)^2+\frac{\mu(Z) \rho(r)}{2}+2 W_3(Z).
\end{eqnarray}
Also for this system the integral $X_1$ reduces to a first order one 
\begin{equation}
\tilde{X}_1= p_\phi^A+\frac{\rho(r)-r^2 \mu(Z)}{2}
\end{equation}
since the magnetic field and the potential are invariant with respect to rotations around $z$-axis. We notice that the system~\eref{p0t0} together with its integrals is actually a limit of~\eref{Wrztau0sol} as $\rho(r)\rightarrow 0$, i.e. as an integrable system~\eref{p0t0} doesn't need to be considered separately.
\end{enumerate}
\item  $\mu(Z)=0$, $\tau(\phi) \neq 0$, $\rho(r)=0$\\
In this case we have $\sigma(r)W_Z=0$, so {\it a priori} there are two possible subcases. However, $\sigma(r)=0$ implies equations of the form~(\ref{appeq1}--\ref{appeq4}) but with $\mu(Z)=0$. Equation~\eref{appeq4} together with our assumptions imposes $W_Z=0$. Thus we must have $W_Z=0$ and the only remaining equation reads
\begin{equation}\label{Wrzmu0}
r^3 \tau'(\phi) \sigma'(r) + 2 \tau'(\phi)\tau(\phi) - 4r^5 W_{r \phi} - 8r^4 W_\phi = 0,
\end{equation}
which is easily integrated. We find 
\begin{eqnarray}\label{Wrzmu0sol}
W &= W_1(r) - \frac{1}{8r^4}\tau(\phi)^2 + \frac{1}{4r^2}\tau(\phi)\sigma(r) + \frac{1}{r^2}W_2(\phi),\\ \nonumber 
B^r &= \frac{1}{2r^2}\tau'(\phi) , \quad B^\phi = \frac{1}{r^3}\tau(\phi) + \frac{1}{2}\sigma'(r), \quad B^Z = 0,
\end{eqnarray}
where $\tau(\phi)$ and $\sigma(r)$ are arbitrary functions, $\tau(\phi)$ not vanishing identically. The integrals~\eref{cylintegrals} are defined by
\begin{eqnarray}\label{Wrzmu0solints}
\fl \nonumber s_1^r & = 0, \, s_1^\phi  = 0, \,  s_1^Z  = \tau(\phi), \,
m_1  = \frac{\tau(\phi)}{2} \left(\sigma(r)-\frac{\tau(\phi)}{r^2}\right)+2 W_2(\phi), \\
\fl s_2^r & = 0, \, s_2^\phi  = 0, \, s_2^Z  = \sigma(r)-\frac{\tau(\phi)}{r^2},\,
m_2  = \frac{1}{4} \left( \sigma(r)-\frac{\tau(\phi)}{r^2} \right)^2.
\end{eqnarray}
The integral $X_2$ can be reduced to a first order one,
\begin{equation}
\tilde{X}_2= p_Z^A+\frac{1}{2} \left( \sigma(r)-\frac{\tau(\phi)}{r^2} \right).
\end{equation}
In cartesian coordinates we have
\begin{eqnarray}
\fl W &= W_1 \left( \sqrt{x^2 + y^2} \right) - \frac{\tau(\phi)^2}{8 \left( x^2+y^2 \right)^2} + \frac{\tau(\phi)\sigma \left( \sqrt{x^2+y^2} \right)}{4(x^2+y^2)} + \frac{W_2(\phi)}{x^2+y^2} , \nonumber \\
\fl B^x &= \frac{x \tau'(\phi)}{2 \left( x^2+y^2 \right)^2} - y \Bigg( \frac{\tau(\phi)}{\left( x^2+y^2 \right)^2} + S\left(\sqrt{x^2+y^2}\right) \Bigg), \nonumber \\
\fl B^y &= \frac{y \tau'(\phi)}{2 \left( x^2+y^2 \right)^2} + x \Bigg( \frac{\tau(\phi)}{\left( x^2+y^2 \right)^2} + S\left(\sqrt{x^2+y^2}\right) \Bigg) , \\
\fl B^z &= 0, \nonumber
\end{eqnarray}\\
where $S(r)=\frac{\sigma'(r)}{2r}$, and $\phi=\arcsin\left( \frac{y}{\sqrt{x^2+y^2}} \right)$.
\item  $\mu(Z)=0$, $\tau(\phi) = 0$, $\sigma(r)=0$ and $\rho(r)\neq 0$

We have $\rho(r)W_\phi=0$, which implies that $W_\phi=0$. Again there is only one equation left to solve 
\begin{equation}
W_{r Z} = 0,
\end{equation}
i.e. we have
\begin{eqnarray}\label{r12dpolar}
W = W_1(r) + W_3(Z), \,
B^r = 0 , \, B^\phi =0 , \, B^Z =  - \frac{1}{2}\rho'(r).
\end{eqnarray}
Thus this class of systems is equivalent to the polar case in two dimensions, which was explored in Ref. \cite{mcsween_2000}, complemented by one-dimensional independent motion in the $z$-direction, governed by the potential $W_3(z)$. The integral $X_2$ becomes the component of the Hamiltonian governing the dynamics in the $z$--direction, the integral $X_1$ is the ``polar'' integral in the $xy$--plane.
\item  $\mu(Z)=0$, $\tau(\phi) = 0$, $\rho(r)=0$, $\sigma(r)\neq 0$. 

We see that $\sigma(r) W_Z=0$ thus $W_Z=0$. There is one remaining equation
\begin{equation}
4r^5 W_{r \phi} - 8r^4 W_\phi = 0
\end{equation}
which is identical to~\eref{Wrzmu0} with $\tau(\phi)=0$. Thus the solution is \begin{eqnarray}\label{Wmtr0sn0}
\fl  W = W_1(r)  + \frac{1}{r^2}W_2(\phi),\quad 
B^r = 0 , \quad B^\phi = \frac{1}{2}\sigma'(r), \quad B^Z = 0
\end{eqnarray}
and the integrals are obtained by setting $\tau(\phi)=0$ in~\eref{Wrzmu0solints}.
\end{enumerate}

\subsection{Case 3b: $\mu(Z) = 0$, $\psi'(\phi) \neq 0$}

Let's recall the reduced row echelon form of $M$ for this case reads~\eref{redMcase2b}. For its rank to be $1$, the only possibility is that both $\sigma(r)$ and $\tau(\phi)$ vanish. Equations~\eref{reducedB} imply that the potential separates as
\begin{equation}\label{r1mun0W}
W(r,\phi,Z)=W_{12}(r,\phi)+W_3(Z).
\end{equation}
Equations~(\ref{reducedAa}--\ref{matrixform2}) reduce to the two following equations which are identical to the ones considered in~\eref{mcsween}:
\begin{eqnarray} \label{rhopsisys}
r\psi'(\phi)W_r + \left( r \rho(r) - \psi(\phi) \right)W_\phi = 0, \nonumber \\
\psi'(\phi) \left( -3 \psi''(\phi) + r^3 \rho''(r) - r^2 \rho'(r) + r \rho(r) - 4 \psi(\phi) \right) \\
+ \psi'''(\phi) \left( r \rho(r) - \psi(\phi) \right) - 4 r^5 W_{r \phi} - 8 r^4 W_\phi = 0. \nonumber
\end{eqnarray}
The magnetic field reads
\begin{equation}\label{r1mun0B}
B^r = 0, \quad B^\phi = 0, \quad B^Z = -\frac{1}{2 r^2} \left( \rho'(r) r^2+\psi''(\phi +\psi(\phi) \right). 
\end{equation}
Thus this class of systems is equivalent to the polar case in two dimensions, which was explored in previous work \cite{mcsween_2000}, complemented by one-dimensional independent motion in the $z$-direction, governed by the potential $W_3(z)$. As above, the integral $X_2$ becomes the component of the Hamiltonian governing the dynamics in the $z$--direction, the integral $X_1$ is the ``polar'' integral in the $xy$--plane.

\section{Conclusions}

Let us first of all sum up the results of this study. The problem stated in the title and Introduction was formulated mathematically in Section~\ref{Himcc} and lead to the determining equations~\eref{cyl2a}--\eref{extra0} for the scalar potential~$W$, the magnetic field~$\vec B$ and the coefficients $\vec s_1$, $\vec s_2$, $m_1$ and $m_2$ of two second order integrals of motion $X_1$ and $X_2$~\eref{IntCyl}. All of the above functions are assumed to be smooth functions of 3 variables, the cylindrical coordinates $r,\phi,Z$ in $\mathbb{E}_3$, with $0\leq r <\infty$, $0\leq \phi< 2 \pi$, $-\infty<Z<\infty$. In Section~\ref{Psderfov} we partially solve this overdetermined system of $28$ PDEs for $12$ functions. We express the vector functions $\vec B$, $\vec s_1$, $\vec s_2$ in terms of 5 scalar auxiliary functions of one variable $\rho(r)$, $\sigma(	r)$, $\tau(\phi)$, $\psi(\phi)$ and $\mu(Z)$, cf.~(\ref{scond}--\ref{Bcond}). We also derive a system of 12 equations~\eref{1_0cond}--\eref{0_0cond} for the remaining scalar functions $m_1$, $m_2$ and $W$ and the auxiliary functions. Some compatibility equations are presented in~\eref{compcond2}.

The reduced system of the determining equations is presented in Section~\ref{reddetsys}. It consists of 3 PDEs for the scalar potential $W(r,\phi,Z)$~\eref{reducedB}, 2 ODEs~(\ref{reducedAa}--\ref{reducedAb}) for the auxiliary functions and 3 algebraic equations~\eref{matrixform2} for the first derivatives $W_r$, $W_\phi$ and $W_z$. Equation~\eref{matrixform2} involves a matrix $M$ depending only on the auxiliary functions. The rank of $M$, $r(M)=r$ satisfies $0\leq r \leq 3$. The case $r=0$ can be discarded since it implies that the magnetic field is absent, $\vec B =0$. In Section~\ref{SdeM3} we show that the reduced determining system has no solutions for $r=3$, i.e. the system is inconsistent.

The main results of this paper are obtained for $r=2$ and $r=3$, presented in Sections~\ref{SdeM2} and~\ref{SdeM1}. The obtained integrable magnetic fields $\vec B(r,\phi,Z)$ and $W(r,\phi,Z)$ are as follows:
\begin{enumerate}
\item $r=2$

The matrix $M$ depends on all $5$ auxiliary functions. The rank condition $r(M)=2$ forces at least one of them to vanish. Three subcases can occur and in all of them the scalar potential splits into two parts as in~\eref{W123}.
\begin{enumerate}
\item $\psi(\phi)=0$

The magnetic field and potential are given in~\eref{rank2psi0cyl}, so $W$ does not depend on $Z$. The second order integrals $X_1$ and $X_2$ are actually squares of first order ones given in~\eref{rank2psi0cyl1int}. They were already found and analysed in an earlier article~\cite{Marchesiello_2015}.
\item $\psi'(\phi)\neq 0$, $\mu(Z)=0$, $(\frac{\tau'(\phi)}{\psi'(\phi)})'=0$

We again find $W_Z=0$ and $\vec B$ is given in~\eref{r2mu0B}. One of the integrals of motion can be reduced to $X_2=p_2$. We obtain a two--dimensional case in~$\mathbb{E}_2$, analyzed earlier in~\cite{mcsween_2000} and~\cite{Berube_2004}. In the perpendicular direction $Z$ we have free motion.
\item $\psi'(\phi)\neq 0$, $\mu(Z)=0$, $(\frac{\tau'(\phi)}{\psi'(\phi)})'\neq 0$

Our analysis leads to the magnetic field~\eref{12B} and the potential~\eref{12W}. Both are expressed in terms of one function $\beta(\phi)=\sqrt{\gamma(\phi)}$ where~$\gamma(\phi)$ satisfies the nonlinear ODE~\eref{gammaeq}.
\end{enumerate} 
\item $r=1$

All 5 auxiliary functions are \emph{a priori} present in $M$ but the rank condition forces at least 2 of them to vanish. Again we obtain several cases:
\begin{enumerate}
\item $\psi(\phi)=\sigma(r)=\rho(r)=0$, $\mu(Z)\neq 0$

The field $\vec B$ and the potential $W$ are expressed in terms of elliptic functions~(\ref{WeiEqs}--\ref{WeiEqsRootsMsol2}). In special cases this simplifies to elementary functions as in~\eref{WeiEqsRootsMsol3}, \eref{WeiEqsRootsMsol4}, \eref{p0C0B} and \eref{p0t0}.
\item $\psi(\phi)=\sigma(r)=\tau(\phi)=0$, $\mu(Z)\neq 0$, $\rho(r)\neq 0$

The result is given in~\eref{Wrztau0sol}.
\item $\psi(\phi)=\mu(Z)=\rho(r)=0$, $\tau(\phi)\neq 0$

We obtain~\eref{Wrzmu0sol}.
\item $\psi(\phi)=\mu(Z)=\sigma(r)=\tau(\phi)=0$, $\rho(r)\neq 0$

This leads to~\eref{r12dpolar}, a case already treated in~\cite{mcsween_2000} for $\mathbb{E}_2$. The motion in the perpendicular $Z$ direction depends on an arbitrary potential $W_3(Z)$.

\item $\psi(\phi)=\tau(\phi)=\mu(Z)=\rho(r)=0$, $\sigma(r)\neq 0$

See~\eref{Wmtr0sn0}.

\item $\psi(\phi)\neq 0$ $\tau(\phi)=\mu(Z)=\sigma(r)=0$

See~\eref{r1mun0W} and~\eref{r1mun0B}. This case again decomposes into integrable motion in $\mathbb{E}_2$ plus a motion governed by $W_3(Z)$ in the perpendicular direction.
\end{enumerate}
\end{enumerate}

This sums up the results on integrable systems of the considered type. Some of the potentials and magnetic fields depend on arbitrary functions as well as constants. This leaves us with the freedom to impose further restrictions, in particular to request that the system be superintegrable, i.e. for 1 or 2 more integrals to exist.

Let us review the differences and similarities between the cases with and
without magnetic fields:
\begin{enumerate}
\item In both cases the leading part of the integral of motion lies in the
enveloping algebra of the Euclidean Lie algebra $\mathfrak{e}_3$.
\item For $\vec B=0$ a second order integral contains no first order terms. For
$\vec B\neq 0$ first order terms can be present.
\item Second order integrability in $\mathbb{E}_n$ implies separation of variables in
the Hamiton--Jacobi and Schr\"odinger equations for $\vec B=0$, but not for $\vec B\neq 0$.
\item For $\vec B=0$ second order integrable and superintegrable systems are the same in quantum and classical mechanics. For $\vec B\neq 0$ this is not necessarily true.
\item For $\vec B\neq \vec 0$ ``exotic potentials'' (expressed in terms of functions satisfying nonlinear ODEs) appear for second order integrability. For $\vec B=\vec 0$ they only appear for higher order integrability and superintegrability ($N\geq 3$). These exotic potentials are typically expressed in terms of elliptic functions, Painlevé transcendents or, in the classical case, solutions of algebraic equations.
\end{enumerate}
   Thus second order integrable and superintegrable systems in magnetic fields are similar to systems without magnetic fields but with integrals of order $N$, $N\geq 3$  \cite{GraWi,Gravel,TreWin,PoWi,MarSajWin,MarSajWin2,EscLVWin1,EscWinYur,EscLVWin2}.

Our future plans include the following. To find all superintegrable systems among the integrable ones in this article. To solve the classical equations of motion and verify that in maximally superintegrable systems all bounded trajectories are closed \cite{Nekhoroshev,TreTurWin1,TreTurWin2}. To determine the cylindrical type quantum integrable and superintegrable systems in a magnetic field. We expect the quantum maximally superintegrable systems to be exactly solvable \cite{TTW,RuhTur,PatWin,Turbiner,Turbiner2}.

\section*{Acknowledgements}

F.F. acknowledges a fellowship from the FESP, Université de Montréal and financial support from the Czech Technical University in Prague during a research visit there. The research of L.\v S. was supported by the Czech Science Foundation (GA\v CR) project 17-11805S. That of P.W. was partially supported by a Discovery grant from NSERC of Canada. The research was largely performed during mutual visits of the members of the author team and we thank each other's institutions for their hospitality.\\

\section*{References}

\bibliographystyle{unsrt}
\bibliography{refs2}

\end{document}